%% file: mains.tex
\documentclass[a4paper,UKenglish,cleveref, autoref, thm-restate]{lipics-v2021}
\usepackage{graphicx}
\usepackage{hyperref}
\usepackage{subfiles}
\usepackage{amsmath}
\usepackage{cleveref}
\usepackage{dsfont}
\usepackage{ifsym}
\usepackage{comment}
\usepackage[dvipsnames]{xcolor}
\usepackage{boxedminipage}
\usepackage[ruled,vlined]{algorithm2e}
\usepackage[nocompress]{cite}
\usepackage{lineno}

\newcommand{\new}{\text{new}}

\DeclareMathOperator{\Ram}{Ram}
\DeclareMathOperator{\col}{col}

\RestyleAlgo{ruled}

\SetKwComment{Comment}{/* }{ */}

\title{Exact Matching: Correct Parity and FPT Parameterized by Independence Number} 

\author{Nicolas {El Maalouly}}{Department of Computer Science, ETH Z\"{u}rich, Switzerland }{nicolas.elmaalouly@inf.ethz.ch}{0000-0002-1037-0203}{}
\author{Raphael {Steiner}}{Department of Computer Science, ETH Z\"{u}rich, Switzerland }{raphaelmario.steiner@inf.ethz.ch}{0000-0002-4234-6136}{Supported by an ETH Zurich Postdoctoral Fellowship.}
\author{Lasse {Wulf}}{Institute of Discrete Mathematics, TU Graz, Austria }{wulf@math.tugraz.at}{0000-0001-7139-4092}{Supported by the Austrian Science Fund (FWF): W1230.}

\authorrunning{N. El Maalouly, R. Steiner, L. Wulf} 

\Copyright{Nicolas El Maalouly, Raphael Steiner, Lasse Wulf} 



\nolinenumbers 

\EventEditors{Satoru Iwata and Naonori Kakimura}
\EventNoEds{2}
\EventLongTitle{34th International Symposium on Algorithms and Computation (ISAAC 2023)}
\EventShortTitle{ISAAC 2023}
\EventAcronym{ISAAC}
\EventYear{2023}
\EventDate{December 3-6, 2023}
\EventLocation{Kyoto, Japan}
\EventLogo{}
\SeriesVolume{283}
\ArticleNo{47}

\begin{document}
\maketitle
\nolinenumbers

\begin{abstract}
Given an integer $k$ and a graph where every edge is colored either red or blue, the goal of the exact matching problem is to find a perfect matching with the property that exactly $k$ of its edges are red. Soon after Papadimitriou and Yannakakis (JACM 1982) introduced the problem, a randomized polynomial-time algorithm solving the problem was described by Mulmuley et al. (Combinatorica 1987). Despite a lot of effort, it is still not known today whether a deterministic polynomial-time algorithm exists. This makes the exact matching problem an important candidate to test the popular conjecture that the complexity classes $\textsc{P}$ and $\textsc{RP}$ are equal. In a recent article (MFCS 2022), progress was made towards this goal by showing that for bipartite graphs of bounded bipartite independence number, a polynomial time algorithm exists. In terms of parameterized complexity, this algorithm was an XP-algorithm parameterized by the bipartite independence number. In this article, we introduce novel algorithmic techniques that allow us to obtain an FPT-algorithm. If the input is a general graph we show that one can at least compute a perfect matching $M$ which has the correct number of red edges modulo 2, in polynomial time. This is motivated by our last result, in which we prove that an FPT algorithm for general graphs, parameterized by the independence number, reduces to the problem of finding in polynomial time a perfect matching $M$ with at most $k$ red edges and the correct number of red edges modulo 2.
\end{abstract}

\subfile{Introduction}
\subfile{Preliminaries}

\subfile{Algorithms}

\subfile{Conclusion}

\bibliography{references}

\appendix
\subfile{Appendix}

\end{document}

%% file: Introduction.tex
\section{Introduction}


In the \emph{Exact Matching Problem} (denoted from now on by EM), we are given a graph $G$ together with a fixed coloring of its edges in two colors (red and blue). 
The question is, for a given integer $k$, to decide whether there exists a perfect matching $M$ of $G$ with the additional property that exactly $k$ of the edges of the perfect matching $M$ are red. Clearly, if we have the special case that all edges of the graph are red and $k= |V(G)| / 2$ then this problem is simply to decide whether there exists a perfect matching in the graph, which is well-known to be decidable in polynomial time \cite{edmonds}. However, when the coloring of the edges is heterogeneous, the problem difficulty seems to increase significantly (see below).

Papadimitriou and Yannakakis~\cite{papadimitriou1982complexity} initially introduced EM in 1982 and conjectured it to be NP-hard. However, a randomized polynomial-time algorithm solving the problem was described by Mulmuley, Vazirani and Vazirani in 1987 in the course of their celebrated isolation lemma~\cite{mulmuley1987matching}. Given standard complexity theoretic hypotheses, this makes it unlikely for EM to be NP-hard. Despite the existence of a polynomial-time randomized algorithm, as of today it is still not known whether EM can also be solved in deterministic polynomial time. The algorithm of Mulmuley et al.\ uses polynomial identity testing and is based on the Schwartz-Zippel Lemma~\cite{schwartz,zippel}, which has resisted all attempts of derandomization so far. Indeed, EM is one of the few natural problems which has a randomized polynomial-time algorithm (i.e.\ it is contained in the complexity class \textsc{\textsc{RP}}) but for which it is not known whether it admits a deterministic polynomial-time algorithm (i.e.\ it is contained in $\textsc{P}$). It is a major open conjecture that \textsc{RP}=\textsc{P}, and so EM becomes a natural candidate to test this hypothesis.

For this reason, EM has been cited in several papers as an open problem. This includes recent breakthrough papers such as the seminal work on the parallel computation complexity of the matching problem~\cite{svensson2017matching}, works on planarizing gadgets for perfect matchings~\cite{gurjar2012planarizing}, works on budgeted, color bounded, or constrained matching problems~\cite{berger2011budgeted,mastrolilli2012constrained,mastrolilli2014bi,stamoulis2014approximation,kelk2019integrality}, on multicriteria optimization problems~\cite{grandoni2010optimization} and on matroid intersection problems~\cite{camerini1992random}.
It is further known that several different problems relate directly or indirectly to EM. The following is a non-exhaustive list of examples: EM is polynomial-time equivalent to the DNA sequencing problem~\cite{blazewicz2007polynomial}. EM is equivalent to a variant of the problem of finding a solution of a binary linear equation system with small Hamming weight \cite{arvind2016solving}. EM can be reduced to a special case of the recoverable robust assignment problem \cite{fischer2020investigation}.

\textbf{Previous work.}
Progress in finding deterministic algorithms for EM (and therefore finding positive evidence for the conjecture P=RP) has only been made for restricted graph classes: It is known that EM can be solved in determinisitic polynomial time for planar and more generally $K_{3,3}$-minor free graphs~\cite{yuster2012almost}, as well as graphs of bounded genus~\cite{genus}. These works use Pfaffian orientations to derandomize the algebraic technique from \cite{mulmuley1987matching}. EM can also be solved for graphs of bounded treewidth using a dynamic programming approach \cite{vardi2022quantum,elmaalouly2022exacttopk}. In contrast to these classes of sparse graphs, EM on dense graphs seems to be even harder: Already solving the problem on complete graphs and complete bipartite graphs is highly nontrivial. In fact, at least 4 articles just dealing with this special case have appeared in the literature \cite{karzanov1987maximum,yi2002matchings,geerdes,gurjar2017exact}. 
Recent work~\cite{elmaalouly2022exact} made a step forward by showing how to solve EM on graphs of constant independence number, where the \emph{independence number} of a graph $G$ is defined as the largest number $\alpha$ such that $G$ contains an \emph{independent set} of size $\alpha$, and bipartite graphs of constant bipartite independence number, where the \emph{bipartite independence number} of a bipartite graph $G$ equipped with a bipartition of its vertices is defined as the largest number $\beta$ such that $G$ contains a \emph{balanced independent set} of size $2\beta$, i.e., an independent set using exactly $\beta$ vertices from both color classes. This generalizes previous results for complete and complete bipartite graphs which correspond to the special cases $\alpha=1$ and $\beta=0$. The authors presented an XP-algorithm, i.e.\ an algorithm running in time $O(n^{f(\alpha)})$, for the problem. The existence of an FPT algorithm, i.e.\ an algorithm with running time $f(\alpha)n^{O(1)}$, was left as an open question. The authors also conjectured that counting perfect matching is \#P-hard for this class of graphs. This conjecture was later proven in \cite{elmaalouly2022hard} already for $\alpha = 2$ or $\beta = 3$. As a consequence, the Pfaffian derandomization technique is unlikely to work for this class of graphs, because this technique implicitly counts the number of perfect matchings. This makes the graph class of graphs of bounded independence number a promising frontier to push the limits of deterministic techniques.
To the best of our knowledge, these are the only results from the last 40 years showing that EM can be solved in deterministic poly-time for restricted graph classes.

 Apart from restricted graph classes, one can also consider parameterized algorithms for EM, using the natural parameter $k$. Note that an XP-algorithm in this case is trivial to obtain using brute-force guessing (guess the red edges that go in a solution and complete the perfect matching using only blue edges). An FPT algorithm would, however, be highly desirable as it is likely provide a lot of insight into EM. The only progress towards that goal can be found in \cite{elmaalouly2022exacttopk} where some color coding tools were developed but only applied to the almost trivial case of bounded circumference graphs. 

 Another direction of progress towards solving EM is the study of relaxed versions of it. A first such relaxation would be to lift the requirement for a perfect matching. In \cite{yuster2012almost}, however, it was shown that there is a simple deterministic polynomial time algorithm such that given a "Yes" instance of EM, computes an \emph{almost} perfect matching (i.e. of size at least $\frac{n}{2}-1$) containing $k$ red edges. This result is as close to optimal as possible for this type of relaxation. The study of the other type of relaxation, i.e. relaxing the color constraints, was only recently initiated in \cite{elmaalouly2022exacttopk}. It was shown that there is a deterministic polynomial time algorithm which given a "Yes"-instance of EM outputs a perfect matching with $k'$ red edges, such that $k/2 \leq k' \leq 3k/2$. 

\textbf{Exact matchings modulo 2.} A crucial tool in this paper is to consider matchings with $k'$ red edges, where $k' \equiv_2 k$, that is, matchings of the correct parity. Let $r(M)$ denote the number of red edges in a matching $M$. We define the \emph{Correct Parity Matching Problem} (CPM), where given a red-blue edge-colored graph and an integer $k$, the goal is to find a perfect matching $M$ such that $r(M) \equiv_2 k$. Note that parity problems (and more general congruency-constrained problems) have been studied in the context of other graph algorithms \cite{hemaspaandra2004complexity,nagele2019submodular}, but are not well studied for the perfect matching problem. A more challenging version of the problem, \emph{Bounded Correct Parity Matching} (BCPM), requires finding a perfect matching $M$ such that $r(M) \equiv_2 k$ \emph{and} $r(M) \leq k$. In \cite{jia2023exact} the complexity of the EM problem was even further highlighted by showing that the Exact Matching polytope has exponential extension complexity even when restricted to the bipartite case and to the parity constraint (i.e. CPM in bipartite graphs has exponential extension complexity).

\textbf{Our results.} 
From now on, when we say that an algorithm has \textit{polynomial running time}, we mean a deterministic algorithm, whose running time is bounded by $\text{poly}(n)$, even if $\alpha, \beta, k$ are not bounded by a constant. We show in this paper how BCPM can be used to solve EM. Precisely, our results are the following:
\begin{itemize}
    \item We show that EM reduces to BCPM in FPT time parameterized by $\alpha$ in the following sense: There exists an algorithm, which performs a single oracle call to BCPM, and solves EM on general graphs, in running time $f(\alpha)n^{O(1)}$. (The result holds analogously for the bipartite independence number $\beta$). Without access to the BCPM oracle, the algorithm outputs a perfect matching with either $k-1$ or $k$ red edges or deduces that the answer of the given EM-instance is "No" (\Cref{sec:reduction-EM-BCPM}).
     \item CPM can be solved in polynomial time for all graphs (\Cref{sec:correct-parity-general}). This insight is based on a deep result by Lov\'{a}sz~\cite{lovasz}. 
    \item On bipartite graphs, the more difficult problem BCPM can be solved in polynomial time (\Cref{th:FPTBCPM}). As a consequence, there is an FPT algorithm parameterized by $\beta$ which solves EM on bipartite graphs (\Cref{thm:main-thm-beta}).
\end{itemize}

Proofs of statements marked $\star$ can be found in the Appendix.

%% file: Preliminaries.tex
\section{Preliminaries}\label{sec:Prel}
\label{sec:prelim}



All graphs considered are simple. For $G = (V, E)$, we let $V(G) := V$ and $E(G) :=  E$. We always use the letter $n$ to denote the number of vertices of a graph $G$, i.e.\ $n = |V(G)|$.  An \textit{edge-colored} graph is a tuple $(G, \col)$, where $\col : E \rightarrow \{\text{red}, \text{blue}\}$ prescribes a color to each edge. An instance of EM is a tuple $(G,\col,k)$. Given an instance of EM and a perfect matching (abbreviated PM) $M$, we define edge weights $w_M: E \rightarrow \mathbb{N}$ as follows: We have $w_M(e) = 0$ if $e$ is a blue edge, $w_M(e) = +1$ if $e$ is a red non-matching edge and $w_M(e) = -1$ if $e$ is a red matching edge. The weight function $w_M$ plays a critical role in many arguments in this paper. When the PM $M$ can be deduced from context, we may write $w$ instead of $w_M$. In this case, the \emph{weight} of edge $e$ is $w_M(e)$.
For $G'$ a subgraph of $G$, we define $R(G')$ (resp. $B(G')$) to be the set of red (resp. blue) edges in $G'$, $r(G') := |R(G')|$ to be the number of red edges of $G'$ and $w_M(G')$ to be the sum of the weights of edges in $G'$. If $\mathcal{C}$ is a set of vertex-disjoint cycles, then we define $w_M(\mathcal{C}) = \sum_{C \in \mathcal{C}}w_M(C)$.

We say that a set of disjoint cycles or paths is $M$-alternating if for any two adjacent edges in the set, one of them is in $M$ and the other is not.
Undirected cycles are considered to have an arbitrary orientation. For a cycle $C$ and $u,v \in C$, $C[u,v]$ is defined as the path from $u$ to $v$ along $C$ (in the fixed but arbitrarily chosen orientation). 
The term $\Ram(r,s)$ refers to the Ramsey number, i.e.\ every graph on $\Ram(r, s)$ vertices contains either a clique of size $r$ or an independent set of size $s$. For simplicity we will use the following upper bound: $\Ram(s+1, s+1) < 4^s$ \cite{ramsey-theory-book}.

%% file: Algorithms.tex
\section{Reducing EM to BCPM in FPT time}
\label{sec:reduction-EM-BCPM}


The goal of this section is to prove our two main theorems:
\begin{theorem}
\label{thm:main-thm-alpha}
EM can be reduced to BCPM in FPT time parameterized by the independence number of the graph.
\end{theorem}
\begin{theorem}
\label{thm:main-thm-beta}
There exists an FPT algorithm for EM on bipartite graphs parameterized by the bipartite independence number of the graph.
\end{theorem}

We will first introduce the algorithm and then prove \Cref{thm:main-thm-alpha} in \Cref{subsec:main-thm-alpha} and \Cref{thm:main-thm-beta} in \Cref{subsec:main-thm-beta}. Finally, in \Cref{subsec:main-thm-no-oracle} we will discuss the case where a BCPM oracle can not be used.



\subsection{Tools from Prior Work}\label{sec:tools}

The algorithm we develop to prove \Cref{thm:main-thm-alpha,thm:main-thm-beta} will rely on many of the tools developed in \cite{elmaalouly2022exact} and \cite{elmaalouly2022exacttopk}. 
We start with the two main propositions that we aim to use.
The setting of both propositions is the same: We are given some PM $M$ explicitly, and we know that there is another PM $M'$ which we know exists, but we do not know explicitly. We are given the PM $M$ and the number $r(M')$ as input and would like to find either $M'$ itself, or at least another PM $M''$ with $r(M'') = r(M')$.

\begin{proposition}[from \cite{elmaalouly2022exact}] \label{prop:smallsetedges}
Let $M$ and $M'$ be two PMs in $G$ such that $|B(M \Delta M')| \leq L$ or $|R(M \Delta M')| \leq L$, for $L \geq 1$. Then there exists a deterministic algorithm running in time $n^{O(L)}$ such that given $M$ and $r(M')$, it outputs a PM $M''$ with $r(M'') = r(M')$. 
\end{proposition}

\begin{proposition}[adapted from \cite{elmaalouly2022exacttopk}]$(\star)$\label{prop:smallsetcycles}
Given a graph $G = (V,E)$ with edge colors red and blue, let $M$ and $M'$ be two PMs in $G$ such that $|E(M \Delta M')| \leq L$, for $L \geq 1$. Then there exists an algorithm running in time $f(L)n^{O(1)}$ (for $f(L) = L^{O(L)}$) such that given $M$ and $r(M')$, it outputs a PM $M''$ with $r(M'') = r(M')$. 
\end{proposition}

The algorithm from \Cref{prop:smallsetcycles} is faster (FPT instead of XP when parameterized by $L$), but it requires more assumptions on $M'$. The algorithm from \Cref{prop:smallsetedges} works by guessing which $L$ edges are in $R(M \Delta M')$ (respectively $B(M \Delta M')$) and then checks if the red (blue) edges can be completed to a PM by using only blue (red) edges. The algorithm from \cref{prop:smallsetcycles} works using color-coding technique (see \cite[Chapter 5]{cygan2015parameterized} for more details on color coding).

In \cite{elmaalouly2022exact} the authors show that for graphs of small independence number, one could use \Cref{prop:smallsetedges} to get an XP algorithm (parameterized by the independence number) by bounding either the number of red edges or the number of blue edges in the symmetric difference with a target matching $M'$. Our aim is to show that we can use the stronger \Cref{prop:smallsetcycles} from \cite{elmaalouly2022exacttopk} to get an FPT algorithm, which would require that we bound both color classes (i.e. the entire symmetric difference). This turns out to be much more difficult to achieve and requires novel algorithmic techniques that we describe in the next section. Our algorithm does, however, start by bounding one of the color classes before bounding the second. For that we simply rely on the tools developed in \cite{elmaalouly2022exact} to avoid starting from scratch.
Due to the technicality of some of the used tools, some readers might want to skip the details of the tools from previous work and jump ahead to the next section, only coming back to these definitions and lemmas when needed. 

A crucial concept to understand the tools from prior work is a property of the weight function $w = w_M$ as defined in \Cref{sec:prelim}. Let $M$ and $M'$ be two perfect matchings. It is well-known that the symmetric difference $\mathcal{C} := M \Delta M'$ is a set of edges that forms a vertex-disjoint union of $M$-alternating cycles. An easy observation is now that $w_M(\mathcal{C})$ counts the difference of red edges between $M$ and $M'$, that is, we have $r(M') = r(M) + w_M(\mathcal{C})$. This follows directly from the definition of $w_M$. The second crucial concept is the concept of a \emph{skip}.
\begin{figure}[ht]
    \centering
    \includegraphics[scale=0.85]{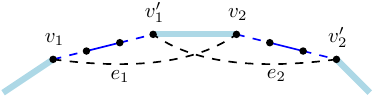}
    \caption{A skip formed by two non-matching edges $e_1$ and $e_2$ (in black). 
    Matching edges are normal lines, non-matching edges are dashed. The bold lines represent subpaths. 
    }
    \label{fig:skip}
\end{figure}
\begin{definition}[from \cite{elmaalouly2022exact}] \label{def:skip}
Let $M$ be a PM and $C$ an $M$-alternating cycle. A skip $S$ is a set of two non-matching edges $e_1 := (v_1,v_2)$ and $e_2 := (v_1',v_2')$ with $e_1, e_2 \notin C$ and $v_1,v_1',v_2,v_2' \in C$ (appearing in this order along $C$) such that $C' = e_1 \cup e_2 \cup C \setminus (C[v_1,v_1'] \cup C[v_2,v_2'])$ is an $M$-alternating cycle, $|C| - |C'| > 0$ and $|w_M(S)| \leq 4$ where $w_M(S):= w_M(C') - w_M(C)$ is called the weight of the skip. 
\end{definition} 

Let $M,C,S,C'$ be as above. We say that \emph{using the skip} $S$ is the action of replacing the alternating cycle $C$ by the alternating cycle $C'$. If furthermore $M'$ is another PM and $C \in M\Delta M'$, then we say that using $S$ also modifies $M'$ the following way: We let $M'_\new = M' \Delta C \Delta C' = M \Delta (((M \Delta M') \setminus C) \cup C')$. In other words, $M'_\new$ is the matching which has the same symmetric difference from $M$ as $M'$, except that $C$ was replaced by $C'$. It is an easy observation that $M'_\new$ is again a PM and $r(M'_\new) = r(M') + w(S)$. This means that using a positive skip (i.e. a skip of strictly positive weight) increases the cycle weight, using a negative skip decreases it and using a \emph{0-skip} (i.e. a skip of weight 0) does not change the cycle weight. Using a skip always results in a cycle of smaller cardinality.
If $P \subseteq C$ is a path and $C[v_1,v_2'] \subseteq P$, then we say that $P$ contains the skip $S$. Two skips $\{(v_1,v_2),(v_1',v_2')\}$ and $\{(u_1,u_2),(u_1',u_2')\}$ are called disjoint if they are contained in disjoint paths along the cycle. Note that two disjoint skips can be used independently. Finally, observe that iterating over all skips of a given alternating cycle $C$ can be done in polynomial time by trying all possible combinations of two chords from the cycle $C$ and checking whether they form a skip. This means that if a skip with certain properties is shown to exist, it can also be found in polynomial time.

\begin{definition}[from \cite{elmaalouly2022exact}]
Let $M$ be a PM and $\mathcal{C}$ a set of disjoint $M$-alternating cycles. A 0-skip set with respect to $\mathcal{C}$ is a set of disjoint skips on cycles of $\mathcal{C}$ such that the total weight of the skips is 0.
\end{definition}

\begin{definition}[from \cite{elmaalouly2022exact}]
Let $M$ be a PM and $\mathcal{C}$ a set of disjoint $M$-alternating cycles. A 0-skip-cycle set with respect to $\mathcal{C}$ is a set of disjoint skips on cycles of $\mathcal{C}$ and/or cycles from $\mathcal{C}$, such that the total weight of the skips minus the total weight of the cycles is 0.
\end{definition}

We say that \emph{using a skip-cycle set} $\mathcal{S}$ means to change $\mathcal{C}$ by removing all cycles in $\mathcal{S}$ from $\mathcal{C}$ and by using all skips in $\mathcal{S}$ (i.e.\ for every $S \in \mathcal{S}$ that is a skip, locate the corresponding cycle $C \in \mathcal{C}$ with $S$ in $C$ and use $S$ on $C$). A perfect matching $M'_\text{new}$ is defined in an analogous way as $M'_\text{new}$ is defined for using a single skip (i.e., such that $M'_\text{new} = M \Delta \mathcal{C}_\text{new}$). 
If $\mathcal{S}$ was a 0-skip-cycle set, then we have $r(M'_\new) = r(M')$. Using a 0-skip-cycle set always decreases the total size of $\mathcal{C}$. In this paper, it will be a common strategy to locate 0-skip-cycle sets contained in the symmetric difference $M\Delta M'$ of two PMs. If we manage to find such a 0-skip-cycle set, then using it on $M\Delta M'$ will produce a new PM $M'_\text{new}$ such that $r(M'_\text{new}) = r(M')$, but $|M\Delta M_\text{new}'| < |M \Delta M'|$. Hence we make progress in the sense that we reduce the symmetric difference $M \Delta M'$ while maintaining $r(M')$.

The following lemmas are taken and adapted from \cite{elmaalouly2022exact}. They show that under certain assumptions 0-skip sets or 0-skip-cycle sets always exist. They are adapted to also include a proof that the desired objects can be found in polynomial time.
We leave their adapted proofs to the appendix.

\begin{lemma}[adapted from \cite{elmaalouly2022exact}]$(\star)$\label{lem:bigcycleskip}
Let $M$ be a PM and $P$ an $M$-alternating path with $w_M(P) \geq 2t\cdot 4^{\alpha}$ (resp. $w_M(P) \leq -2t\cdot 4^{\alpha}$), for $t \geq 1$, then $P$ contains at least $t$ disjoint negative (resp. positive) skips. If $P$ and $M$ are given, then we can also find $t$ such skips in polynomial time. 
\end{lemma}

\begin{lemma}[adapted from \cite{elmaalouly2022exact}]$(\star)$\label{lem:boundedcycleweights}
Let $t \geq 8\cdot4^{\alpha}$ and $t' = 4t^2$.
Let $M$ be a PM and $\mathcal{C}$ a set of disjoint $M$-alternating cycles and $C \in \mathcal{C}$ such that $|w_M(\mathcal{C})| \leq t'$ and $|w_M(C)| \geq 2t'$, then $\mathcal{C}$ contains a 0-skip-cycle set. 
If $\mathcal{C}, M$ are given, we can also find a 0-skip-cycle set in polynomial time.

\end{lemma}

\begin{lemma}[adapted from \cite{elmaalouly2022exact}]$(\star)$\label{lem:notmanycycles} 
Let $t \geq 3$.
Let $M$ be a PM and $\mathcal{C}$ a set of disjoint $M$-alternating cycles such that $|w_M(\mathcal{C})| \leq t$, $|w_M(C)| \leq 2t$ for all $ C \in \mathcal{C}$ and $|\mathcal{C}| \geq 10t^3$, then $\mathcal{C}$ contains a 0-skip-cycle set. If $\mathcal{C}, M$ are given, we can also find a 0-skip-cycle set in polynomial time. 
\end{lemma}


\begin{lemma}[adapted from \cite{elmaalouly2022exact}]$(\star)$\label{lem:notmanyredandblue}
Let $t \geq 8\cdot4^{\alpha}$.
Let $M$ be a PM and $\mathcal{C}$ a set of disjoint $M$-alternating cycles such that $|\mathcal{C}| \leq 10t^3$, $|w_M(C)| \leq 2t$ for all $ C \in \mathcal{C}$ and $\mathcal{C}$ contains at least $1000t^6$ blue edges and $1000t^6$ red edges, then $\mathcal{C}$ contains a 0-skip set. 
If $\mathcal{C}, M$ are given, then we can also find a 0-skip set in polynomial time. 
\end{lemma}

\subsection{The Main Algorithm}\label{sec:algorithm}

The aim of this section is to present the algorithm which reduces EM to BCPM in time $f(\alpha)n^{O(1)}$. We first sketch the idea of the algorithm: We assume that the algorithm receives two PMs $M$ and $M'$ as input, such that $r(M) < k < r(M')$ and such that both $M$ and $M'$ already have the correct parity, i.e.\ $r(M) \equiv_2 r(M') \equiv_2 k$. We will later show how this can be done with an oracle call to BCPM. 
But even in the case where the BCPM oracle can not be used and $M, M'$ are just some arbitrary PMs with $r(M) < k < r(M')$, our algorithm still computes something meaningful: We show that in this case a PM with $k$ or $k-1$ red edges will be output, or it will be deduced that the given EM-instance has answer "No". This variant of the algorithm is further discussed in \Cref{subsec:main-thm-no-oracle}.

Our algorithm modifies the PMs $M$ and $M'$ many times. But the invariant is maintained that during the whole execution of the algorithm, both the PMs $M$ and $M'$ will never change their parity. The basic idea of the algorithm is to have many iterations, where in each iteration either $M$ is modified such that $r(M)$ increases by 2, or $M'$ is modified such that $r(M')$ decreases by two. Clearly, if we can do such a modification in every iteration, we will eventually arrive at a PM with $k$ red edges. One might ask why we consider modifications of the kind $+2$ and $-2$, instead of the kind $+1$ and $-1$.  The reason for this is that a change of $\pm 1$ might not always be possible, even in complete graphs. To see this, consider the smallest possible modification of a PM. It consists in taking its symmetric difference with an alternating cycle of length four. Such a cycle may add or remove up to two red edges from the matching and it is possible that we only find such cycles adding or removing exactly two red edges. On the converse, if all small cycles add or remove one red edge, we can still achieve a change of two by simply considering two such cycles.

However, reality is more complicated and even a $\pm 2$ modification might not always be possible. The first hurdle is that such a modification might not be possible if $r(M) \ll r(M')$. 
To combat this hurdle, the algorithm splits into three phases, where in the first phase the PMs $M,M'$ are modified such that they keep their parity and after their modification we have that $r(M), r(M')$ are close to $k$. Details for phase 1 will be provided in \Cref{th:falphacloseparity}. 
In the second phase, we will do many iterations, such that in each iteration the algorithm tries to (i) increase $r(M)$ by 2, or (ii) decrease $r(M')$ by 2, or (iii) strictly decrease the cardinality of the symmetric difference $|E(M \Delta M')|$. Finally, it can still happen that neither (i), (ii), or (iii) are possible.  However, we prove a key lemma which states that in this situation (and if the given EM-instance is a "Yes" instance), we can use color coding techniques to find a PM $M^*$ in time $f(\alpha)n^{O(1)}$ which is a solution to EM, i.e.\ $r(M^*) = k$. The algorithm then enters phase 3, where it either finds $M^*$ or deduces that the given EM-instance is a "No"-instance. We now provide the formal description of the algorithm: 
\subparagraph{Input:}
A red-blue edge-colored graph $(G, \col)$, a nonnegative integer $k$. Two PMs $M$ and $M'$ with $r(M) < k < r(M')$.
\subparagraph{Phase 1:}
Find two PMs $M_\new$ and $M'_\new$ such that 
\[k-8\cdot 4^\alpha \leq r(M_\new) \leq k \leq r(M'_\new) \leq k+8\cdot 4^\alpha,\] and such that the parity is maintained, i.e.\ $r(M_\new) \equiv_2 r(M)$ and $r(M'_\new) \equiv_2 r(M')$. Set $M \gets M_\new$ and $M' \gets M'_\new$. If this step fails, output "EM-instance has no solution".

\subparagraph{Phase 2:}
If either $M$ or $M'$ is a solution matching we are done. Otherwise repeat the following three steps until either $M$ or $M'$ is a solution matching
or until every step (i),(ii), and (iii) fails in the same iteration:
\begin{enumerate}
    \item[(i)] Invoke the algorithm of \Cref{prop:smallsetedges} with respect to the matching $M$ and $L = 2$ in order to try to find a PM $M_\new$ with $r(M_\new) = r(M) + 2$. If such a PM is found, let $M \gets M_\new$, otherwise do not modify $M$ and consider step (i) as failed.

    \item[(ii)] Invoke the algorithm of \Cref{prop:smallsetedges} with respect to the matching $M'$ and $L = 2$ in order to try to find a PM $M'_\new$ with $r(M'_\new) = r(M') - 2$. If such a PM is found, let $M' \gets M'_\new$, otherwise do not modify $M'$ and consider step (ii) as failed.
    \item[(iii)] Invoke the algorithms of \Cref{lem:boundedcycleweights}, \Cref{lem:notmanycycles} or \Cref{lem:notmanyredandblue} (with $t = 256\cdot 4^{2\alpha}$), to try to find a 0-skip or a 0-skip-cycle set in $M \Delta M'$. If such an object is found, then use it (i.e. change $M'$ accordingly) to reduce $|E(M \Delta M')|$. Otherwise do not modify $M,M'$ and consider step (iii) as failed.
\end{enumerate}

\subparagraph{Phase 3:}
If either $M$ or $M'$ is a solution matching
we are done. Otherwise invoke the algorithm of \Cref{prop:smallsetcycles} with $L= 2^{\alpha^{O(1)}}$ (for appropriately large constants) on the matching $M$ to try to find a PM $M^*$ with $r(M^*) = k$. If such a PM $M^*$ is found, then output it. Otherwise output "EM-instance has no solution".

\subparagraph{}
This completes the description of the algorithm. The remainder of this section is dedicated to its proof.  First, we prove that phase 1 can be completed correctly in polynomial time (\Cref{th:falphacloseparity}). It is not so difficult to prove that phase 2 requires only polynomial time (as there are at most $n^2$ iterations).
Finally, we prove in our main lemma (\Cref{lem:symdiffbound}) that if steps (i),(ii),(iii) all fail simultaneously, then phase 3 is guaranteed to succeed. This is the most difficult lemma to prove. In \Cref{subsec:main-thm-alpha} we summarize the proof and explain how to obtain the two initial matchings $M$ and $M'$ required as input for phase 1. 

Finally, we describe the modifications necessary for bipartite graphs (\Cref{subsec:main-thm-beta}) and for cases where the BCPM oracle is not available (\Cref{subsec:main-thm-no-oracle}).

\subsection{Proof of the main lemmas}
\label{subsec:proof-main-lemmas}



The following lemmas help us prove the correctness and polynomial running time of the algorithm.

\begin{lemma} \label{th:falphacloseparity}
Given a "Yes" instance of EM and two PMs $M$ and $M'$ with $r(M) \leq k \leq r(M')$, there exists a deterministic polynomial time algorithm that outputs two PMs $M_1$ and $M_2$ with $r(M_1) \equiv_2 r(M) $, $r(M_2) \equiv_2 r(M') $ and $k-8\cdot 4^\alpha \leq r(M_1) \leq k \leq r(M_2) \leq k+8\cdot 4^\alpha$.
\end{lemma}

\begin{proof}
As long as $r(M) < k-8\cdot 4^{\alpha}$ we will consider two cases:
\begin{itemize}
    \item All cycles $C \in M \Delta M'$ have weight $w_M(C) \leq 4\cdot 4^\alpha$. In this case $M \Delta M'$ must contain at least two strictly positive cycles $C_1$ and $C_2$. If $ w_M(C_1) \equiv_2 0$ then we replace $M$ by $M \Delta C_1$ and iterate (note that $r(M) < r(M \Delta C_1) \leq k$ and $r(M \Delta C_1) \equiv_2 r(M)$). The case $ w_M(C_2) \equiv_2 0$ is similar. Otherwise we replace $M$ by $M \Delta (C \cup C')$ and iterate (note that $r(M) < r(M \Delta (C \cup C')) \leq k$ and $r(M \Delta (C \cup C')) \equiv_2 r(M)$).
    \item There exists $C \in M \Delta M'$ with $w_M(C) > 4\cdot 4^\alpha$. Observe that $C \in M' \Delta M$ and $w_{M'}(C) = - w_M(C) \leq -4\cdot 4^\alpha$. By \Cref{lem:bigcycleskip} applied to $M'$ and $w_{M'}$, we have that $C$ contains two positive skips (with respect to $M'$ and $w_{M'}$). If any of the skips has even weight, we use it to increase the weight of $w_{M'}(C)$ and iterate (note that $r(M)$ increases since using a skip in $M' \Delta M$ modifies $M$). Otherwise we use both skips. In either case, $r(M)$ increases and its parity is preserved. Note that $r(M)$ can increase by at most 8 given that a skip must have weight at most $4$ by definition.
\end{itemize}
In both cases $r(M)$ increases after every iteration. So there can be at most $O(n)$ iterations, each running in polynomial time, until $ k-8\cdot 4^{\alpha} \leq r(M) \leq k$.
Now we apply a similar procedure to decrease $r(M')$. As long as $r(M') > k+8\cdot 4^{\alpha}$ we will consider two cases:
\begin{itemize}
    \item All cycles in $M' \Delta M$ have weight $w_{M'}$ more than $-4\cdot 4^\alpha$. In this case $M' \Delta M$ must contain at least two strictly negative cycles $C_1$ and $C_2$. If $ w_{M'}(C_1) \equiv_2 0$ then we replace $M'$ by $M' \Delta C_1$ and iterate (note that $k \leq r(M'\Delta C_1) < r(M')$ and $r(M'\Delta C_1) \equiv_2 r(M')$). The case $ w_{M'}(C_2) \equiv_2 0$ is similar. Otherwise we replace $M'$ by $M' \Delta (C \cup C')$ and iterate (note that $k \leq r(M \Delta (C \cup C')) < r(M')$ and $r(M \Delta (C \cup C')) \equiv_2 r(M')$).
    \item There exists $C \in M' \Delta M$ with $w_{M'}(C) < -4\cdot 4^\alpha$. Observe that $C \in M \Delta M'$ with $w_M(C) = -w_{M'}(C) \geq 4\cdot 4^\alpha$. By \Cref{lem:bigcycleskip} applied to $M$ and $w_M$, $C$ contains two negative skips (with respect to $M$ and $w_M$). If any of the skips has even weight, we use it to reduce $w_M(C)$ and iterate (note that $r(M')$ decreases since using a skip in $M \Delta M'$ modifies $M'$). Otherwise we use both skips. In either case, $r(M')$ decreases and its parity is preserved. Note that $r(M')$ can decrease by at most 8 given that a skip must have weight at least $-4$ by definition.
\end{itemize}
In both cases $r(M')$ decreases after every iteration. So there can be at most $O(n)$ iterations, each running in polynomial time, until $ k \leq r(M') \leq k+8\cdot 4^{\alpha}$.
Finally the algorithm terminates by outputting $M_1 := M$ and $M_2 := M'$.
\end{proof} 




\begin{lemma}\label{lem:longmono}
Let $M$ be a PM and $\mathcal{C}$ a set of disjoint $M$-alternating cycles with the following properties:
\begin{itemize}
    \item $\mathcal{C}$ does not contain monochromatic cycles.
    \item $|E(\mathcal{C})| \geq 2t^3$.
    \item $|R(\mathcal{C})| \leq t$ (resp. $|B(\mathcal{C})| \leq t$).
\end{itemize}
Then $\mathcal{C}$ contains a blue (resp. red) $M$-alternating path of length at least $t$.
\end{lemma}

\begin{proof}

We will consider the case when $|R(\mathcal{C})| \leq t$. The case $|B(\mathcal{C})|\leq t$ is proven similarly by swapping the two colors. First observe that if $\mathcal{C}$ contains at most $t$ red edges and no monochromatic cycles, then $|\mathcal{C}|\leq t$. So by the pigeonhole principle, $\mathcal{C}$ must contain a cycle $C$ with $|E(C)| \geq 2t^2$. Consider the set of maximal blue subpaths of $C$ and let $p_B$ be the number of these paths. As every such path is accompanied by a red edge, we have $p_B \leq t$. Finally, $C$ has at least $2t^2 - t$ blue edges, so by the pigeonhole principle one of the blue paths must have length at least $(2t^2 - t)/t \geq t$. 
\end{proof}

The above lemma simply shows that if only one color class is bounded, there must be long monochromatic paths of the other color. The next lemma shows that the existence of long monochromatic paths in turn implies the existence of small cycles.


\begin{lemma}\label{lem:skiporCycle}
Let $M$ be a PM and $C$ an $M$-alternating cycle. Let $P \subseteq C$ be a blue (resp. red) $M$-alternating path
of length at least $6\Ram(\Ram(4,\alpha+1),\alpha+1)$, starting with a non-matching edge and 
not containing 0-skips. Then there must be two edges $e_1 := (b_1,b_2)$ and $e_2 := (w_1,w_2)$ with endpoints on $P$, at least one of which must be red (resp. blue), such that $C' = e_1 \cup e_2 \cup C[b_1,w_1] \cup C[b_2,w_2]$ is an $M$-alternating cycle with $0< w_M(C')\leq 2$ (resp. $-2\leq w_M(C')< 0$) and containing a number of red (resp. blue) edges equal to the absolute value of its weight.
\end{lemma}

\begin{figure}[ht]
    \centering
    \includegraphics[scale=0.85]{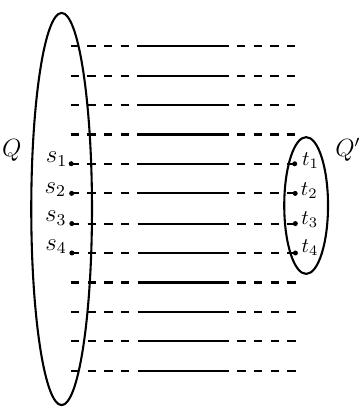}
    \includegraphics[scale=0.85]{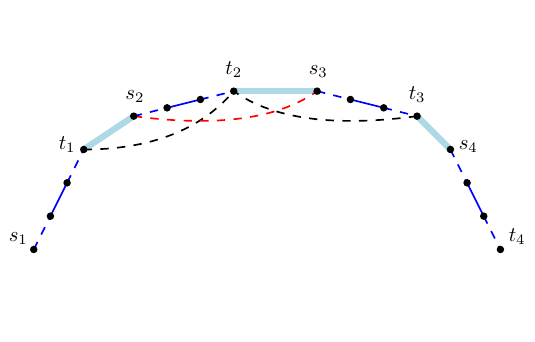}
    \caption{Left: The set of paths $\mathcal{P}_2$, of size $\Ram(4,\alpha + 1)$, and the cliques $Q$ and $Q'$. Right: The paths from $\mathcal{P}_3$ along the blue path $P$ and the vertices $s_i, t_j$. Matching edges are normal lines, non-matching edges are dashed. The bold lines between $t_i$ and $s_{i+1}$ represent subpaths. Observe that $\{(s_2,s_3),(t_2,t_3)\}$ forms a skip and $(t_1,t_2) \cup (s_2,s_3) \cup C[t_1,s_2] \cup C[t_2,s_3] $ is an alternating cycle.}
    \label{fig:skiporcycle}
\end{figure}

\begin{proof}
We will only deal with the case when $P$ is blue, the other case is treated similarly (by switching the roles of the two colors in the proof). We assume that $P$ has an arbitrary orientation which is used to define the start and end vertices of subpaths of $P$. First, we divide $P$ into a set of consecutive paths $\mathcal{P}$ of length 6 each, starting with the first non-matching edge. Let $\mathcal{P}_1$ be the set of paths formed by the first 3 edges of each path in $\mathcal{P}$. The set of start vertices of paths in $\mathcal{P}_1$ has size at least $\Ram(\Ram(4,\alpha+1),\alpha+1)$ so it must contain a clique $Q$ of size $\Ram(4,\alpha+1)$. Let $\mathcal{P}_2$ be the set of paths in $\mathcal{P}_1$ with start vertices in $Q$. The set of end vertices of paths in $\mathcal{P}_2$ must contain a clique $Q'$ of size $4$ (see \Cref{fig:skiporcycle}). Let $\mathcal{P}_3 := \{P_1, P_2, P_3, P_4\}$ be the set of paths in $\mathcal{P}_1$ with end vertices in $Q'$. Let $s_i$ and $t_i$ be the start and end vertices of path $P_i$ for $i\in \{1,2,3,4\} $. Observe that any two distinct paths $P_i, P_j \in \mathcal{P}_3$ have their endpoints connected by the edges $(s_i,s_j)$ and $(t_i,t_j)$ and a skip is created this way. If both edges were blue, we would get a 0-skip (since the whole path $P$ has only blue edges). Letting $i=2,j=3$, we see that one of the edges $(s_2,s_3)$ or $(t_2,t_3)$ must be red. Suppose $(s_2,s_3)$ is red. Observe that $(t_1,t_2) \cup (s_2,s_3) \cup C[t_1,s_2] \cup C[t_2,s_3] $ is a cycle of weight 
$+1$ or $+2$ (depending on whether $(t_1,t_2)$ is red or blue, since all other edges are blue) and containing at most 2 red edges. Similarly, suppose $(t_2,t_3)$ is red. Observe that $(t_2,t_3) \cup (s_3,s_4) \cup  C[t_2,s_3] \cup C[t_3,s_4] $ is a cycle of weight $+1$ or $+2$ (depending on whether $(s_3,s_4)$ is red or blue) and the number of red (resp. blue) edges it contains is equal to the absolute value of its weight.
\end{proof}

Finally, we are ready to prove our main lemma. Roughly speaking, it states that if all steps (i),(ii) and (iii) in phase two of the algorithm fail, then phase 3 is guaranteed to succeed. More specifically, it states that if we cannot make small progress towards a solution then we are ready to apply \Cref{prop:smallsetcycles} and find one in FPT time. Small progress here means either getting the number of red edges in $M$ or $M'$ closer to $k$, or making their symmetric difference smaller.

\begin{lemma}
\label{lem:symdiffbound} 
Let $M$ and $M'$ be two PMs with the following properties:
\begin{itemize}
    \item[(a)] $r(M) < k-1$, $r(M') > k$.
    \item[(b)] $|w_M(M \Delta M')| \leq t$ for $t = 256\cdot 4^{2\alpha}$.
    \item[(c)] There is no PM $M_1$ such that $r(M_1) = r(M) + 2 $ and $|R(M \Delta M_1)| = 2$.
    \item[(d)] There is no PM $M_1'$ such that $r(M_1') = r(M') - 2$ and $|B(M' \Delta M_1')| = 2$.
    \item[(e)] $M \Delta M'$ does not contain any 0-skip.
    \item[(f)] The algorithms of \Cref{lem:boundedcycleweights}, \Cref{lem:notmanycycles} and \Cref{lem:notmanyredandblue} all fail to find a 0-skip-cycle set in $M \Delta M'$.
\end{itemize}
If there is at least one PM with $k$ red edges, then there exists a PM $M^*$ such that $r(M^*) = k$ and $|E(M \Delta M^*)| = 2^{\alpha^{O(1)}}$.
\end{lemma}

\begin{proof}

We start by giving a high level overview and the intuition behind the proof.
First we note that properties (a) and (b) will be guaranteed after phase 1 of the algorithm and they state that both $M$ and $M'$ are close to $k$ in terms of number of red edges. Second, properties (c) and (d) state that our algorithm is unable to make small progress in terms of getting the number of red edges in $M$ or $M'$ closer to $k$. Finally properties (e) and (f) state that our algorithm is unable to make small progress in term of making the symmetric difference between $M$ and $M'$ smaller.

Our final goal is to bound the symmetric difference between $M$ and some solution $M^*$. We will do that by contradiction to one of the given properties. Observe that if the conditions for \Cref{lem:skiporCycle} are met, i.e. there is a long blue only $M$-alternating path, then the lemma guarantees that small progress towards getting the number of red edges closer to $k$ is possible. The same holds for long red only $M'$-alternating paths. Note, however, that we might need to apply the lemma twice in order to ensure that the progress is in increments or decrements of two, thus contradicting either property (c) or (d). 
This way we can bound the length of blue only $M$-alternating paths. Then if red only $M$-alternating paths are also bounded, \Cref{lem:longmono} implies that either both colors are bounded, in which case we are done, or none of them is. In the latter case, we use the machinery developed in \cite{elmaalouly2022exact} (see \Cref{sec:tools}) to reach the contradiction (remember that the goal there was to bound one color class), and this requires properties (a) and (b) to hold. The same holds if blue only $M'$-alternating paths are bounded.

The only remaining obstacles are long red only $M$-alternating or blue only $M'$-alternating paths. To deal with that, we try to reduce the symmetric difference between $M$ and $M'$ such that long monochromatic $M$-alternating paths are also $M'$-alternating (and the contradiction above can again be reached) since the two matchings do not differ by that many edges. Bounding the symmetric difference between $M$ and $M'$ relies on a contradiction to properties (e) or (f). It follows the same steps as bounding the symmetric difference between $M$ and $M^*$, but with the added benefit that paths in this symmetric difference are both $M$ and $M'$-alternating, which avoids the problem of long red only $M$-alternating or blue only $M'$-alternating paths.

To summarise, we start by bounding $|E(M \Delta M')|$ (first bounding one color class, then the second). Then we are able to bound $|E(M \Delta M^*)|$ (again one color class at a time).

\subparagraph{Detailed proof.}
We will start by showing that one color class of $M \Delta M'$ must be bounded. This allows us to then bound $|E(M \Delta M')|$. We then consider the solution matching $M^*$ that minimizes $|E(M \Delta M^*)|$ and start by bounding the number of blue edges in $M \Delta M^*$. Finally, we also show that the number of red edges in $M \Delta M^*$ is bounded, thus bounding $|E(M \Delta M^*)|$.

\subparagraph{Bounding one color class of $M \Delta M'$.}
Since we failed to reduce $|E(M \Delta M')|$ using the algorithm of \Cref{lem:boundedcycleweights}, the weight of all cycles in $M \Delta M'$ must be bounded: $|w(C)| \leq 2t$ for all $ C \in M \Delta M'$.
Since we failed to reduce $|E(M \Delta M')|$ using the algorithm of \Cref{lem:notmanycycles}, the number of cycles in $M \Delta M'$ must be bounded: $|M \Delta M'| \leq 10t^3$.
Finally, since we failed to reduce $|E(M \Delta M')|$ using the algorithm of \Cref{lem:notmanyredandblue}, $|B(M \Delta M')|$ or $|R(M \Delta M')|$ must be bounded (by $1000t^6$). Let $t' = \max(1000t^6, 20\Ram(\Ram(4,\alpha+1),\alpha+1))$ (note that $1000t^6 = 2^{O(\alpha)}$, $\Ram(4,\alpha+1) = \alpha^{O(1)}$ and $\Ram(\Ram(4,\alpha+1),\alpha+1)) = 2^{\alpha^{O(1)}}$ so $t' = 2^{\alpha^{O(1)}}$).

\subparagraph{Bounding $|E(M \Delta M')|$.}

First, we show that property (c) implies that there is no blue $M$-alternating path of length at least $t'$ in the graph. Suppose such a path exists. 
Divide the path into two blue paths $P_1$ and $P_2$ of length at least $t'/2$ each. From \Cref{lem:skiporCycle} applied to each of the paths $P_1$ and $P_2$, we get that there exists two disjoint $M$-alternating cycles $C_1$ and $C_2$ with $0<w_M(C_1)\leq 2$, $0<w_M(C_2)\leq 2$ and each containing a number of red edges equal to the absolute value of their weight. If $C_1$ contains two red edges, let $M_1 := M \Delta C_1$. Otherwise if $C_2$ contains two red edges, let $M_1 := M \Delta C_2$. Finally, if both $C_1$ and $C_2$ contain only one red edge, let $M_1 := M \Delta (C_1 \cup C_2)$. Observe that $|R(M \Delta M_1)| = 2$ and $r(M_1) = r(M) + 2 $, contradicting property (c).

Next we show that property (d) implies that there is no red $M'$-alternating path of length at least $t'$ in the graph.
Divide the path into two red paths $P_1$ and $P_2$ of length at least $t'/2$ each. From \Cref{lem:skiporCycle} applied to each of the paths $P_1$ and $P_2$, we get that there exists two disjoint $M$-alternating (with respect to $M'$) cycles $C_1$ and $C_2$ with $-2\leq w_{M'}(C_1)< 0$, $-2\leq w_{M'}(C_2)< 0$ and and each containing a number of blue edges equal to the absolute value of their weight. If $C_1$ contains two blue edges, let $M_1' := M' \Delta C_1$. Otherwise if $C_2$ contains two blue edges, let $M_1' := M' \Delta C_2$. Finally, if both $C_1$ and $C_2$ contain only one blue edge, let $M_1' := M' \Delta (C_1 \cup C_2)$. Observe that $|B(M' \Delta M_1')| = 2$ and $r(M_1') = r(M') - 2 $, contradicting property (d).

Suppose $|R(M \Delta M')| \geq 2t'^3$. Then by the previous paragraph $|B(M \Delta M')|\leq t'$. Note that $M \Delta M'$ contains no monochromatic cyle, as this would be a 0-skip cycle set, therefore by \Cref{lem:longmono}, $M' \Delta M$ contains a red $M'$-alternating path of length at least $t'$. But this contradicts property (c).
Now suppose $|B(M \Delta M')| \geq 2t'^3$. Then $|R(M \Delta M')|\leq t'$ and by \Cref{lem:longmono}, $M \Delta M'$ contains a blue $M$-alternating path of length at least $t'$, contradicting property (c). So we get $|E(M \Delta M')| = |B(M \Delta M')| + |B(M \Delta M')| \leq 4t'^3$.

\subparagraph{Bounding $|B(M \Delta M^*)|$.}
Now let $M^*$ among all those PMs with $k$ red edges be the one which minimizes $|E(M \Delta M^*)|$. Note that $|w_M(M\Delta M^*)| \leq |w_M(M\Delta M')| \leq t$.
Since $|E(M \Delta M^*)|$ is minimal, $M \Delta M^*$ cannot contain a 0-skip-cycle set. By \Cref{lem:boundedcycleweights}, the weight of all cycles in $M \Delta M^*$ must be bounded: $|w(C)| \leq 2t$ for all $ C \in M \Delta M^*$.
By \Cref{lem:notmanycycles}, the number of cycles in $M \Delta M^*$ must be bounded: $|M \Delta M^*| \leq 10t^3$.
Finally, by \Cref{lem:notmanyredandblue}, $|B(M \Delta M^*)|$ or $|R(M \Delta M^*)|$ must be bounded (by $1000t^6 \leq t'$). Suppose $|B(M \Delta M^*)| > 2t'^3$. So $|R(M \Delta M^*)| \leq t'$ and by \Cref{lem:longmono}, $M \Delta M^*$ contains a blue $M$-alternating path of length $t'$, contradicting property (c). So $|B(M \Delta M^*)| \leq 2t'^3$.

\subparagraph{Bounding $|E(M \Delta M^*)|$.}

Let $t'' = 4t'^4$.
Suppose $|R(M \Delta M^*)| > 2t''^3$, by \Cref{lem:longmono} $M^* \Delta M$ contains a red path $P$ with $|P| \geq t''$.
Observe that $M' \Delta M^* = (M \Delta M^*) \Delta (M \Delta M')$ and $P \subseteq M \Delta M^*$ so
$$P \cap (M' \Delta M^*) = P \backslash (P \cap (M \Delta M')). $$
We have $|P \cap (M \Delta M')| \leq |E(M \Delta M')| < 4t'^3$, so if all paths in $P \cap (M' \Delta M^*)$ have length at most $t'$ then $|P| < 4t'^4$, a contradiction. So there must be a path $P' \subseteq P \cap (M' \Delta M^*)$ of length at least $t'$. Note that $P'$ is a red $M'$-alternating path, contradicting property (d).
So we have $|E(M \Delta M^*)| \leq 4t''^{3} = t''^{O(1)} = t'^{O(1)} = 2^{\alpha^{O(1)}}$.
\end{proof}

\subsection{Main theorem for general graphs}
\label{subsec:main-thm-alpha}

\begin{proof}[Proof of \Cref{thm:main-thm-alpha}]
Suppose we have a polynomial time oracle for BCPM.
We start by solving CPM on the given instance. This can be done in polynomial time (as we prove later in \Cref{th:FPTCPM}) and will give us a PM $M_p$ with $r(M_p) \equiv_2 k$. 
If $r(M_p) \geq k$, let $M' := M_p$ and use an oracle call to BCPM to get $M$ with $r(M) \equiv_2 k$ and $r(M) \leq k$. 
Otherwise, if $r(M_p) \geq k$, let $M := M_p$ and use an oracle call to BCPM to get $M'$ with $r(M') \equiv_2 k$ and $r(M') \geq k$ (this can simply be done by swapping the red and blue colors and using $k'=n/2 -k$ as parameter for the BCPM oracle). In both cases, we obtain PMs $M,M'$ such that $r(M) \equiv_2 k \equiv_2 r(M')$ and $r(M) \leq k \leq r(M')$.

Note that if this step fails (in the sense that the CPM or BCPM call returns "false"), then the EM instance has no solution. Otherwise we apply the algorithm of \Cref{sec:algorithm} on the EM-instance with $M$ and $M'$ as input. Our goal now is to prove that if the EM-instance is a "YES" instance, then the following must be true:
\begin{itemize}
    \item[(a)] Phase 1 runs in polynomial time and outputs two PMs $M$ and $M'$ such that $k - 8 \cdot 4^\alpha \leq r(M) \leq k \leq r(M') \leq k + 8 \cdot 4^ \alpha$, $r(M) \equiv_2 r(M') \equiv_2 k$.
    \item[(b)] Phase 2 runs in polynomial time and either outputs a PM with $k$ red edges (and the algorithm terminates) or a PM $M$ such that there exists a PM $M^*$ with $r(M^*) = k$ and $|E(M \Delta M^*)| \leq 2^{\alpha^{O(1)}}$ (for appropriately large constants).
    \item[(c)] If the algorithm did not terminate in Phase 2, then Phase 3 runs in time $f(\alpha) n^{O(1)}$ and outputs a PM with $k$ red edges.
\end{itemize}

It is easy to see that if all the above items hold, then the algorithm runs in time $f(\alpha)n^{O(1)}$ and always outputs a PM with $k$ red edges if one exists. Note that (a) and (c) follow directly from \Cref{th:falphacloseparity} and \Cref{prop:smallsetcycles} respectively.

To prove (b) first observe that as long as $r(M) \neq k$ and $r(M') \neq k$, all steps in phase 2 maintain the following invariants: $r(M) \leq k \leq r(M')$ and $r(M) \equiv_2 r(M') \equiv_2 k$. To see this, note that $r(M)$ and $r(M')$ can only change by $2$ every step and they start with the same parity as $k$. So in order for $r(M)$ to go above $k$ or $r(M')$ to go below $k$ they would need to pass by $k$, at which point the algorithm terminates. Also observe that if any of the steps does not fail, then either $r(M') - r(M)$ decreases or $|E(M \Delta M')|$ decreases while $r(M') - r(M)$ remains unchanged. So if we consider as a measure of progress $r(M') - r(M)$ and $|E(M \Delta M')|$ ordered lexicographically (where progress is towards smaller values of the measure), then we always make progress (i.e. the measure strictly decreases). Note that $r(M') - r(M) \leq n$ and is always non-negative and the same holds for $|E(M \Delta M')|$. So the algorithm can perform at most $n^2$ iterations in phase 2. Since every iteration runs in polynomial time (this is true for steps (i) and (ii) by \Cref{prop:smallsetedges} and for step (iii) by \Cref{lem:boundedcycleweights}, \Cref{lem:notmanycycles} and \Cref{lem:notmanyredandblue}), we get that phase 2 runs in polynomial time.
Now observe that the algorithm only terminates in phase 2 if either $M$ or $M'$ is a solution (i.e. it has $k$ red edges). So it remains to show that if the algorithm does not terminate in this phase then there exists a PM $M^*$ with $r(M^*) = k$ and $|E(M \Delta M^*)| \leq 2^{\alpha^{O(1)}}$. Observe that in case of non-termination, all the conditions of \Cref{lem:symdiffbound} are met: 
\begin{itemize}
    \item[(a)] $r(M) < k-1$, $r(M') > k$: follows from the invariants and $M$, $M'$ not being solutions.
    \item[(b)] $|w_M(M \Delta M')| \leq 256\cdot 4^{2\alpha}$: follows from $r(M') - r(M) \leq  16\cdot 4^{\alpha}$.
    \item[(c)] There is no PM $M_1$ such that $r(M_1) = r(M) + 2 $ and $|R(M \Delta M_1)| = 2$: follows from the failure of (i).
    \item[(d)] There is no PM $M_1'$ such that $r(M_1') = r(M') - 2$ and $|B(M' \Delta M_1')| = 2$: follows from the failure of (ii).
    \item[(e)] $M \Delta M'$ does not contain any 0-skip: follows from the failure of (iii).
    \item[(f)] The algorithms of \Cref{lem:boundedcycleweights}, \Cref{lem:notmanycycles} and \Cref{lem:notmanyredandblue} all fail to find a 0-skip-cycle set in $M \Delta M'$: follows from the failure of (iii).
\end{itemize}
So by \Cref{lem:symdiffbound} we get the desired result.
\end{proof}

\subsection{Main theorem for bipartite graphs}
\label{subsec:main-thm-beta}



In order to prove the main theorem for the bipartite case (\Cref{thm:main-thm-beta}), we start by proving a similar result to the main theorem on general graph that is adapted to bipartite graphs, i.e., we use the bipartite independence number of the graph (the proof can be found in the Appendix).

\begin{lemma}$(\star)$
\label{lem:main-thm-beta}
EM on bipartite graphs can be reduced to BCPM on bipartite graphs in FPT time parameterized by the bipartite independence number of the graph.
\end{lemma}

It remains to show that there is a deterministic polynomial time algorithm for BCPM on bipartite graphs. This result can be derived from the more general result of \cite{artmann2017strongly} on network matrices, as noted in \cite{jia2023exact}, even for the more general weighted version of the problem. To make it more accessible, we reprove it using a standard dynamic programming techniques. The high level approach, as briefly described in \cite{jia2023exact}, is the following: start by computing a minimum weight perfect matching, in our case a perfect matching with minimum number of red edges, and if the number of red edges is even then find a minimum odd weight alternating cycle and output the symmetric difference. We could not find formal proof of correctness and running time for this algorithm in the literature, therefore we provide one in the Appendix.

\begin{theorem}$(\star)$\label{th:FPTBCPM}
There is a deterministic polynomial time algorithm for BCPM on bipartite graphs.
\end{theorem}

\subsection{Main theorem without oracle access}
\label{subsec:main-thm-no-oracle}

Although an FPT algorithm parameterized by the independence number for general graphs still requires an algorithm for BCPM, the following theorem shows that without relying on BCPM the algorithm developed in this section can still output a PM that is very close to optimal, i.e. it contains either $k$ or $k-1$ red edges (the proof is similar to that of \Cref{thm:main-thm-alpha} and left for the Appendix). 

\begin{theorem}$(\star)$
\label{thm:approx-thm-alpha}
There exists an algorithm such that given a "Yes" instance of EM, it outputs a perfect matching with either $k-1$ or $k$ red edges in time $f(\alpha)n^{O(1)}$.
\end{theorem}

This strengthens the results of \cite{elmaalouly2022exacttopk} by reducing the constraint violation to at most one red edge at the expense of an FPT (parameterized by the independence number) instead of a polynomial running time.


\section{Correct Parity Matching for General Graphs}
\label{sec:correct-parity-general}

While solving BCPM for general graphs remains an open problem, in this section we present a solution to the easier problem of CPM which is only concerned with the parity of the number of red edges.

\begin{theorem}$(\star)$\label{th:FPTCPM}
There is a deterministic polynomial time algorithm for CPM.
\end{theorem}

We will establish Theorem~\ref{th:FPTCPM} as a consequence of a deep result by Lov\'{a}sz~\cite{lovasz} on the \emph{linear hull} of perfect matchings of a graph. We first need to introduce some notation which we adopt from~\cite{lovasz}. Let a (not necessarily bipartite) graph $G=(V,E)$ and a field $\mathbb{F}$ be given, and let us denote by $\mathcal{M}$ the set of perfect matchings of $G$. Then the linear hull of perfect matchings $\text{lin}_{\mathbb{F}}(\mathcal{M})$ is the linear subspace of $\mathbb{F}^E$, generated by the characteristic vectors of perfect matchings in $G$. Concretely, $\text{lin}_{\mathbb{F}}(\mathcal{M})$ is the linear span of $\{1_M|M \in \mathcal{M}\}$, where for every perfect matching $M$ the vector $1_M \in \mathbb{F}^E$ is defined by $1_M(e)=1$ for every $e\in M$ and $1_M(e)=0$ for every $e \in E\setminus M$. 

We will make use of the following result of Lov\'{a}sz~\cite{lovasz}.

\begin{theorem}[\cite{lovasz}]\label{th:lovasz}
For every finite field $\mathbb{F}$ there is a deterministic polynomial-time algorithm that, given as input a graph $G$, returns a linear basis of $\text{lin}_{\mathbb{F}}(\mathcal{M})$.
\end{theorem}

The importance of this result by Lov\'{a}sz for solving the CPM is explained through the following lemma. As usual, for two vectors $x, y \in \mathbb{F}^E$ we denote by $\langle x,y\rangle :=\sum_{e \in E}{x_ey_e} \in \mathbb{F}$ their scalar product. 

\begin{lemma}
\label{lemma:oddmatchings}
Let $G=(V,E)$ be a graph equipped with a coloring of its edges with colors red and blue. Let $\mathbb{F}_2$ denote the $2$-element field and let $\{x_1,\ldots,x_d\} \subset \mathbb{F}_2^E$ be a linear basis of $\text{lin}_{\mathbb{F}_2}(\mathcal{M})$. Let $r \in \mathbb{F}_2^E$ be defined by $r_e:=1$ for all red edges $e \in E$ and $r_e=0$ for all blue edges $e \in E$. Then the following two statements are equivalent:
\begin{enumerate}
    \item There exists a perfect matching $M$ in $G$ containing an odd number of red edges.
    \item There exists $i \in \{1,\ldots,d\}$ such that $\langle x_i, r\rangle=1$.
\end{enumerate}
\end{lemma}

\begin{proof}
Suppose first that there exists a perfect matching $M$ in $G$ containing an odd number of red edges. Then the incidence vector $1_M \in \text{lin}_{\mathbb{F}_2}(\mathcal{M})$ can be represented as a linear combination of the basis elements $x_1, \ldots, x_d$, in other words, there exists $I \subseteq \{1,\ldots,d\}$ such that 
$$1_M=\sum_{i \in I}{x_i}.$$
Taking scalar products with $r$ we get
$$\langle 1_M, r \rangle=\sum_{i \in I}{\langle x_i, r\rangle}.$$ Note that the scalar product on the left hand side equals the number of red edges in $M$ taken modulo $2$, and hence it equals $1$. But then at least one of the scalar products on the right hand side must also be non-zero, i.e., there exists $i \in I$ with $\langle x_i, r \rangle=1$, as desired.

Conversely, suppose there exists $i \in \{1,\ldots,d\}$ such that $\langle x_i, r\rangle=1$. Then by virtue of $\text{lin}_{\mathbb{F}_2}(\mathcal{M})$ being spanned by the characteristic vectors of perfect matchings in $M$, there exists a list of perfect matchings $M_1,\ldots,M_t$ in $G$ such that $x_i=\sum_{j=1}^{t}{1_{M_j}}$. Using the same argument as above, i.e., by taking scalar products with $r$ and using that $\langle x_i, r\rangle=1$, we find that there must exist $j \in \{1,\ldots,t\}$ with $\langle 1_{M_j},r\rangle=1$, which means that $M_j$ is a perfect matching of $G$ with an odd number of red edges. This concludes the proof. 
\end{proof}

We may now deduce the following.

\begin{corollary}\label{cor:decisions}
There exists a deterministic polynomial-time algorithm, that, given as input a red-blue edge-colored graph $G=(V,E)$ and a number $k \in \mathbb{Z}$, decides whether or not $G$ contains a perfect matching $M$ with $r(M) \equiv_2 k$. 
\end{corollary}
\begin{proof}
Suppose first that $k$ is odd. We use Theorem~\ref{th:lovasz} to compute in deterministic polynomial time a linear basis $x_1,\ldots,x_d$ of $\text{lin}_{\mathbb{F}_2}(\mathcal{M})$. Note that since $\text{lin}_{\mathbb{F}_2}(\mathcal{M})$ is a subspace of $\mathbb{F}_2^E$, its dimension satisfies $d \le |E|$. Next we generate the incidence vector $r$ of red edges as in the previous lemma, and compute the scalar products $\langle x_i, r\rangle$ for $i=1,\ldots,d$ in polynomial time. If at least one of these products equals $1$, we return that a perfect matching $M$ with $r(M) \equiv_2 k$ exists, and otherwise we return that such a matching does not exist. The correctness of this output follows by Lemma~\ref{lemma:oddmatchings}.

Next suppose $k$ is even. Let $G'$ be the red, blue-edge colored graph which is obtained as the disjoint union of $G$ with its given edge-coloring and a disjoint new edge of color red. The perfect matchings of $G'$ are exactly the perfect matchings of $G$ together with the additional new red edge, and hence $G$ contains a perfect matching $M$ with $r(M)\equiv_2 k$ if and only if $G'$ contains a perfect matching with and odd number of red edges. Thus we can decide whether such a matching exists by invoking the algorithm from the case $k=1$ described above with $G'$ as the input. 
\end{proof}

It is now easy to use the above decision-version of the CPM to solve the CPM itself by a standard edge-deletion procedure.
\begin{proof}[Proof of Theorem~\ref{th:FPTCPM}]
Let $G=(V,E)$ be the input graph with a given red, blue-edge coloring, and let further $k \in \mathbb{Z}$ be given. We use Corollary~\ref{cor:decisions} to decide if $G$ contains a perfect matching $M$ with $r(M)\equiv_2 k$. If it does not, then the algorithm stops with this conclusion. Otherwise, we search through the edges $e \in E$ one by one, and for each such edge test (again using Corollary~\ref{cor:decisions}) whether $G-e$ contains a perfect matching $M$ with $r(M)\equiv_2 k$.

Suppose first we find an edge $e \in E$ such that $G-e$ contains a perfect matching $M$ with $r(M)\equiv_2 k$. In this case we make a recursive call of the algorithm to $G-e$, which will return a perfect matching with the correct parity in $G-e$. We can then return this matching, as it is also a perfect matching in $G$ with the correct parity of red edges. 

Otherwise, we find that there exists no $e \in E$ such that $G-e$ contains a perfect matching $M$ with $r(M)\equiv_2 k$. But as we know that $G$ does contain a perfect matching $M$ with $r(M)\equiv_2 k$, this means that all edges of $G$ are contained in $M$, and hence we may return the set of edges of $G$ and thereby find a solution to the CPM. 
\end{proof}


%% file: Conclusion.tex
\section{Conclusion and Open Problems}\label{sec:conc}

So far, EM has only been solved for very sparse graphs (i.e.\ bounded tree-width and bounded genus graphs) and very dense graphs (i.e.\ bounded independence number graphs). The techniques used are quite different between these two cases. Especially in the case of dense graphs, many previous works considered only complete (bipartite) graphs without much progress. Only recently, the results were extended to the case of bounded independence number, leading to XP algorithms parameterized by $\alpha$ or $\beta$. Looking for FPT algorithms was the natural next step. In this paper, we could resolve the bipartite case fully, while the non-bipartite case could only be resolved partially. However, our results in the non-bipartite case still yield the following two non-trivial, independent insights: (i) To obtain an FPT algorithm parameterized by $\alpha$ it suffices to solve BCPM (\Cref{thm:main-thm-alpha}) and the easier problem CPM can always be solved (\Cref{th:FPTCPM}). (ii) An FPT algorithm parameterized by $\alpha$ can w.l.o.g. assume to start with a PM with $k-1$ red edges (\Cref{thm:approx-thm-alpha}).

We hope that these insights can be the starting point for future work to obtain an FPT algorithm parameterized by $\alpha$, or, even better, an FPT algorithm parameterized by $k$. The latter would be considered quite a breakthrough as it is likely to require a lot of deep understanding of the structure and patterns behind EM, given the difficulty of making progress towards it.


%% file: Appendix.tex
\section{Missing Proofs from \Cref{sec:reduction-EM-BCPM}}

\subsection{Tools from Prior Work}

\begin{proof}[Proof of \Cref{prop:smallsetcycles}]
The proof is the same as in \cite{elmaalouly2022exacttopk} but using a uniform weight function on the edges since the algorithm there requires a weighted graph as input.
\end{proof}

To adapt the proofs of \Cref{lem:boundedcycleweights}, \Cref{lem:notmanycycles} and \Cref{lem:notmanyredandblue} we need some more definitions and lemmas from \cite{elmaalouly2022exact}.

\begin{lemma}[adapted from \cite{elmaalouly2022exact}]\label{lem:skipfrompaths}
Let $M$ be a PM and $P$ an $M$-alternating path containing a set $\mathcal{P}$ of disjoint paths, each of length at least $3$ and starting and ending at non-matching edges, of size $|\mathcal{P}| \geq 4^{\alpha}$. Then $P$ contains a skip.
If all paths in $\mathcal{P}$ have the same weight $x$, then if $x$ is one of the following values, we get the following types of skips:
\begin{itemize}
    \item $x=2$: negative skip.
    \item $x=1$: negative or 0-skip.
    \item $x=0$: positive or 0-skip.
    \item $x=-1$: positive skip.
\end{itemize}
If $P$ is given, we can find such a skip in $P$ in polynomial time.
\end{lemma}

\begin{proof}
The set of starting vertices of the paths in $\mathcal{P}$ must contain a clique $Q$ of size $\alpha + 1$ since $|\mathcal{P}| > \Ram(\alpha+1, \alpha+1)$ (and the independence number of the graph is $\alpha$). Let $\mathcal{P}'$ be the set of paths from $\mathcal{P}$ starting with vertices in $Q$ and $Q'$ their set of ending vertices. Since $|Q'| = \alpha + 1$, there must be an edge connecting two of its vertices, call it $e_2$. Let $P_1$ and $ P_2$ be the two paths in $\mathcal{P}'$ connected by $e_2$. Let $e_1$ be the edge connecting the starting vertices of $P_1$ and $P_2$ (which must exist since $Q$ is a clique). Note that $e_1$ and $e_2$ must be non-matching edges. Now observe that $e_1$ and $e_2$ form a skip $S$ and $w(S) = w(e_1) + w(e_2) - w(P_1) - w(P_2)$. Finally, suppose $P_1$ and $P_2$ have weight $x$ and note that $w(e_1),w(e_2) \in \{0,1\}$ since they are non-matching edges. We get $-2x \leq w(S) \leq 2-2x$ thus proving the lemma.

If $P$ is given, finding such a skip can be done in polynomial time by simply guessing its two edges (i.e. trying all combinations of two edges with endpoints on $P$ and checking if they form the desired skip).
\end{proof}

\begin{lemma}[adapted from \cite{elmaalouly2022exact}]\label{lem:0skipset}
Let $\mathcal{S}$ be a collection of disjoint skips. If $\mathcal{S}$ contains at least 4 positive skips and at least 4 negative skips (all mutually disjoint), then $\mathcal{S}$ must contain a 0-skip set. 
If $\mathcal{S}$ is given, such a 0-skip set can be computed in constant time.
\end{lemma}

\begin{proof}
Note that all considered skips have weight (in absolute value) in the set $\{1,2,3,4\}$. The lemma can be simply proven by enumerating all possibilities for the positive and negative skips.
Also note that finding such a 0-skip set can be done by enumerating all possible subsets of skips (of which there are constantly many), each time checking if the total weight of the subset is 0.
\end{proof}

\begin{definition}[from \cite{elmaalouly2022exact}]
A $+1$ pair (resp. $-1$ pair and $0$ pair) is a pair of consecutive edges (the first a matching-edge and the second a non-matching edge) along an $M$-alternating cycle such that their weight sums to $1$ (resp. $-1$ and $0$).
\end{definition}

Two $+1$ (resp. $-1$) pairs are called consecutive if there is an $M$-alternating path between them on the cycle which only contains 0 pairs.
\begin{definition}[from \cite{elmaalouly2022exact}]
A $+1$ (resp. $-1$) bundle is a pair of edge-disjoint consecutive $+1$ (resp. $-1$) pairs. The path starting at the first pair and ending at the second one is referred to as the bundle path.
\end{definition}

\begin{definition}[from \cite{elmaalouly2022exact}]
A Sign Alternating Path (SAP) is an $M$-alternating path $P$ formed by edge pairs, such that it does not contain any bundles.
\end{definition}

\begin{lemma}[from \cite{elmaalouly2022exact}]\label{lem:manyzeros}
Let $M$ be a PM and $P$ an $M$-alternating path containing at least $10t^3$ blue (resp. red) edges. Then one of the following properties must hold:
\begin{itemize}
    \item[(a)] $P$ contains at least $t$ disjoint bundles.
    \item[(b)] $P$ contains an SAP with at least $t$ non-zero pairs.
    \item[(c)] $P$ contains at least $t$ edge-disjoint 0-paths of length at least 4 starting with a blue (resp. red) matching edge.
\end{itemize}
\end{lemma}

\begin{lemma}[from \cite{elmaalouly2022exact}]\label{lem:manypaths}
A path $P$, satisfying one of the following properties, must contain $t$ disjoint paths each of length at least $3$, starting and ending with non-matching edges and having specific weights that depend on the satisfied property:
\begin{itemize}
    \item[(a)] $P$ contains $t$ disjoint $+1$ bundles: paths of weight $+2$.
    \item[(b)] $P$ contains $t$ disjoint $-1$ bundles: paths of weight $-1$.
    \item[(c)] $P$ contains $t$ edge-disjoint 0-paths of length at least 4 starting with a red matching edge: paths of weight $+1$.
    \item[(d)] $P$ contains $t$ edge-disjoint 0-paths of length at least 4 starting with a blue matching edge: paths of weight $0$.
    \item[(e)] $P$ contains an SAP with at least $2t+1$ non-zero pairs: paths of weight $+1$.
    \item[(f)] $P$ contains an SAP with at least $2t+1$ non-zero pairs: paths of weight $0$.
\end{itemize}

\end{lemma}



\begin{proof}[Proof of \Cref{lem:bigcycleskip}]
We will only prove the case $w(P) \geq 2t\cdot 4^{\alpha}$, the case $w(P) \leq -2t\cdot 4^{\alpha}$ is proven similarly. First we prove that $P$ must contain at least $t\cdot 4^{\alpha}$ disjoint $+1$ bundles. Suppose not. Let $\mathcal{B}$ be a maximum size set of disjoint $+1$ bundles in $P$. Let $P'$ be the path obtained from $P$ by contracting the bundle paths of bundles in $\mathcal{B}$. Observe that $w(P') \geq w(P) - 2(t\cdot 4^{\alpha}-1) \geq 2$. 

We claim that $P'$ cannot contain any $+1$ bundles. Suppose there exists such a bundle $B$ in $P'$. $P$ cannot contain $B$ (by maximality of $\mathcal{B}$), so $B$'s non-zero pairs are separated in $P$ by a set of $+1$ bundles $\mathcal{B}' \subseteq \mathcal{B}$ (otherwise it would still be contained in $B$). Now consider the path $P'' \subseteq P$ starting and ending with $B$'s non-zero pairs. Let $\mathcal{B}''$ be the set of $+1$ bundles formed by pairs of consecutive $+1$ pairs along $P''$. Observe that $|\mathcal{B}''| \geq |\mathcal{B}'|+1 $ and that $\mathcal{B} \backslash \mathcal{B}' \cup \mathcal{B}'' $ is a set of disjoint $+1$ bundles in $P$ of size larger than $\mathcal{B}$, a contradiction. So $P'$ cannot contain any $+1$ bundles.

Since $P'$ does not contain any consecutive $+1$ pairs, $w(P') \leq 1$ (since every $+1$ pair is now followed by a $-1$ pair), a contradiction. So $P$ must contain at least $t\cdot 4^{\alpha}$ disjoint $+1$ bundles. 
Finally we take a maximum size set of such bundles and group together every $4^{\alpha}$ consecutive ones along $P$ then use \Cref{lem:manypaths} and \Cref{lem:skipfrompaths} to get a negative skip for each group. 

Observe that grouping the bundles (i.e. finding edge-disjoint paths along $P$, each containing one group) can be done in polynomial time (by simply walking along $P$ and counting bundles) and \Cref{lem:skipfrompaths} guarantees that we can then find each of the skips in polynomial time.
\end{proof}



\begin{proof}[Proof of \Cref{lem:boundedcycleweights}]
Suppose $w(C) \geq 2t'$ (the case $w(C) \leq -2t'$ is treated similarly). 
By \Cref{lem:bigcycleskip}, $C$ contains at least $4t$ disjoint negative skips (which can be found in polynomial time), of which a subset $\mathcal{S}$ of size at least $t$ must have the same weight $-w_1$, with $1 \leq w_1 \leq 4$ (by the pigeonhole principle). Note that $\mathcal{S}$ can be computed in polynomial time given the above skips.
Now we consider 2 cases:

Case (1): $\mathcal{C}$ contains a cycle $C'$ with $w(C') \leq -t$. Then by \Cref{lem:bigcycleskip} $C'$ contains at least 4 positive disjoint skips (which can be found in polynomial time), so $\mathcal{C}$ contains a 0-skip set by \Cref{lem:0skipset} (which can be found in polynomial time).

Case (2): $w(C') > -t$, $\forall C' \in \mathcal{C}$. In this case $\mathcal{C}$ contains at least $4t$ negative cycles (otherwise $w(\mathcal{C}) > w(C) - 4t^2 \geq 8t^2 - 4t^2 \geq t'$), so there must be at least $4$ cycles in $\mathcal{C}$ of the same weight $-w_2$ with $1 \leq w_2 \leq t$ (we can find such cycles in polynomial time given $\mathcal{C}$). Observe that $w_1$ of these cycles along with $w_2$ of the skips in $\mathcal{S}$ form a 0-skip-cycle set, which can then be found in polynomial time. 
\end{proof}



\begin{proof}[Proof of \Cref{lem:notmanycycles}]

First note that a cycle of weight 0 is also a 0-skip-cycle set, so we will assume that no such cycle exists in $\mathcal{C}$. Now suppose $\mathcal{C}$ contains at least $4t^2$ positive and $4t^2$ negative cycles. There must be at least $2t$ cycles of same positive weight $w_1 \leq 2t$ and $2t$ cycles of same negative weight $-w_2 \geq -2t$ (we can find such cycles in polynomial time given $\mathcal{C}$). The set of $w_2$ cycles of weight $w_1$ and $w_1$ cycles of weight $-w_2$ is a 0-skip-cycle set which can thus be found in polynomial time. Now assume w.l.o.g. $\mathcal{C}$ contains less than $4t^2$ positive cycles and let $x$ be the number of negative cycles in $\mathcal{C}$. Then $-t \leq w(\mathcal{C}) < 4t^2 \cdot 2t - x = 8t^3 -x$ so $x < 8t^3 + t < 10t^3-4t^2$. But this implies $|\mathcal{C}| < 10t^3$, a contradiction.
\end{proof}


\begin{lemma}[from \cite{elmaalouly2022exact}] \label{lem:minusimpliesplus}
Let $C$ be a cycle with $|w(C)| \leq l$. If $C$ contains $3t+ l$ disjoint $-1$ (resp. $+1$) bundles, then $C$ also contains at least $t$ disjoint $+1$ (resp. $-1$) bundles. 
\end{lemma}


\begin{lemma}[adapted from \cite{elmaalouly2022exact}]\label{lem:notmanybundles}
Let $t \geq 8\cdot 4^{\alpha}$.
Let $C$ be a cycle with $|w(C)| \leq 2t$. If $C$ contains more than $10t$ disjoint bundles then it must contain a 0-skip set. 
If $C$ is given, then we can also find a 0-skip set in polynomial time.
\end{lemma}

\begin{proof}
Suppose $C$ contains less than $2t$ disjoint $+1$ bundles.
By \Cref{lem:minusimpliesplus}, $C$ contains at most $8t$ disjoint $-1$ bundles, a contradiction to the total number of bundles. So $C$ contains at least $2t$ disjoint $+1$ bundles. Similarly, $C$ contains at least $2t$ disjoint $-1$ bundles. 

By \Cref{lem:manypaths}, we get that $C$ contains at least $2t$ subpaths of weight $-1$ and $2t$ subpaths of weight $2$, all edge-disjoint, starting and ending with non-matching edges.
Now we cut $C$ into two paths $P_1$ and $P_2$, such that $P_1$ contains at least $t$ paths of weight $-1$ while $P_2$ still contains at least $t$ paths of weight $+2$ (to see that this works, simply note that by walking along the path and stopping as soon as we have covered $t$ of the subpaths of some weight, we are left with a path that contains at least $t$ subpaths of the other weight).
We divide $P_1$ (resp. $P_2$) into paths each containing at least $2\cdot4^{\alpha}$ of these subpaths (note that there are at least 4 such paths for each of $P_1$ and $P_2$). By \Cref{lem:skipfrompaths} they each contain at least a positive (resp. negative) skip.
Finally by \Cref{lem:0skipset}, $C$ contains a 0-skip set.
Note that such a 0-skip set contains at most 8 skips, i.e. 16 edges. This means that we can find it in polynomial time by simple guessing its edges. 

\end{proof}


\begin{proof}[Proof of \Cref{lem:notmanyredandblue}]
Observe that some cycle $C_1 \in \mathcal{C}$ must contain at least $100t^3$ blue edges and some cycle $C_2 \in \mathcal{C}$ must contain at least $100t^3$ red edges. If $C_1 \neq C_2$ we let $P_1 := C_1$ and $P_2 := C_2$. Otherwise we cut $C_1$ into two paths $P_1$ and $P_2$, such that $P_1$ contains at least $50t^3$ blue edges while $P_2$ still contains at least $50t^3$ red edges (to see that this works, simply note that by walking along the path and stopping as soon as we have covered $50t^3$ of the edges of some color class, we are left with a path that contains at least $50t^3$ edges of the other class).

Now by \Cref{lem:manyzeros} we know that one of the following must be true:
\begin{itemize}
    \item[(a)] $C_1$ (resp. $C_2$) contains at least $10t$ disjoint bundles (if $C_1 \neq C_2$ then \Cref{lem:manyzeros} can be applied to both $P_1$ and $P_2$, otherwise it can be applied to $P_1 \cup P_2$).
    \item[(b)] $P_1$ (resp. $P_2$) contains an SAP with at least $5t$ non-zero pairs.
    \item[(c)] $P_1$ (resp. $P_2$) contains at least $5t$ edge-disjoint 0-paths of length at least 4 starting with a blue (resp. red) matching edge.
\end{itemize}

For case (a) we get a 0-skip set by \Cref{lem:notmanybundles}.
For cases (b) and (c), by \Cref{lem:manypaths}, we get that $P_1$ (resp. $P_2$) contains at least $t$ edge-disjoint subpaths starting and ending with non-matching edges and of weight $0$ (resp. $1$). Suppose $P_1$ (resp. $P_2$) does not contain 0-skips (otherwise we are done). We divide $P_1$ (resp. $P_2$) into 4 paths each containing at least $2\cdot4^{\alpha}$ of these subpaths. By \Cref{lem:skipfrompaths}, they each contain at least one positive (resp. negative) skip.
Finally by \Cref{lem:0skipset}, $\mathcal{C}$ contains a 0-skip set.
Note that such a 0-skip set contains at most 8 skips, i.e. 16 edges. This means that we can find it in polynomial time by simply guessing its edges. 
\end{proof}


\subsection{Main theorem for bipartite graphs}

To prove the above \Cref{lem:main-thm-beta} we need to consider the concept of a \emph{biskip} instead of a skip (see definition below). 
Recall that the tools from prior work used in the non-bipartite case were \Cref{lem:bigcycleskip,lem:boundedcycleweights,lem:notmanycycles,lem:notmanyredandblue}. 
It is proven in \cite{elmaalouly2022exact} that exactly the same lemmas remain true if "$\alpha$" is replaced by "$\beta$" and "skip" is replaced by "biskip". (We adapted \Cref{lem:bigcycleskip,lem:boundedcycleweights,lem:notmanycycles,lem:notmanyredandblue} slightly to additionally show the existence of a poly-time algorithm. It can easily be shown in the same manner that the same adaption can be made in this case.) Note that the proofs of \Cref{th:falphacloseparity,lem:longmono,lem:symdiffbound} do not rely on the actual properties of the independence number $\alpha$. Only the proof of \Cref{lem:skiporCycle} does rely on it. \Cref{lem:skiporCyclebi} provides a bipartite analogue of \Cref{lem:skiporCycle}. Then \Cref{lem:main-thm-beta} is proven by adapting all involved lemmas to use "$\beta$" instead of "$\alpha$" and "biskip" instead of "skip". As this is a straightforward adaption from the previous proof, we omit the details. 

\begin{definition}[adapted from \cite{elmaalouly2022exact}]\label{def:biskip}
Let $M$ be a PM and $C$ an $M$-alternating cycle. A biskip $S$ is a set of two edges $e_1 := (v_1,v_2)$ and $e_2 := (v_1',v_2')$ with $e_1, e_2 \notin C$ and $v_1,v_2',v_1',v_2 \in C$ (appearing in this order along $C$) such that $C_1 := C[v_2,v_1] \cup e_1 $ and $C_2 := C[v_2',v_1'] \cup e_2 $ are vertex disjoint  $M$-alternating cycles, $|C| - |C_1| - |C_2| > 0$ and $|w(S)| \leq 4$ where $w(S) := w(C_1) + w(C_2) - w(C)$ is called the weight of the biskip. A biskip can also be defined by only one edge $e_1$ in which case the cycle $C_2$ is the empty cycle.
\end{definition}

\begin{lemma}\label{lem:skiporCyclebi}
Let $M$ be a PM and $C$ an $M$-alternating cycle in a bipartite graph. Let $P \subseteq C$ be a blue (resp. red) $M$-alternating path
of length at least $6(2\beta+2)^2$, starting with a non-matching edge and 
not containing 0-biskips. Then there must be two edges $e_1 := (b_1,b_2)$ and $e_2 := (w_1,w_2)$ with endpoints on $P$, at least one of which must be red, such that $C' := e_1 \cup e_2 \cup C[b_1,w_2] \cup C[w_1,b_2]$ is an $M$-alternating cycle with $0< w_M(C')\leq 2$ (resp. $-2\leq w_M(C')< 0$) and containing a number of red (resp. blue) edges equal to the absolute value of its weight.
\end{lemma}

\begin{proof}
We will only deal with the case when $P$ is blue, the other case is treated similarly. First, we divide $P$ into a set of consecutive paths $\mathcal{P}$ of length 6 each, starting with the first non-matching edge. Let $\mathcal{P}_1$ be the set of paths formed by the first 3 edges of each path in $\mathcal{P}$. Note that $|\mathcal{P}_1| \geq (2\beta+2)^2$. Let $\mathcal{P}_2 \subset \mathcal{P}_1$ be the 1st, $(2\beta + 2)+1$th, $2(2\beta + 2)+1$th,... paths in $\mathcal{P}_1$. 
Note that $|\mathcal{P}_2| = 2\beta + 2$ and between any two paths in $\mathcal{P}_2$ along $P$ there are $2\beta + 1$ paths from $\mathcal{P}_1$. 
Let $V_1$ be the set of end vertices of the first $\beta + 1$ paths in $\mathcal{P}_2$ and $V_2$ the set of start vertices of the last $\beta + 1$ paths in $\mathcal{P}_2$. Observe that $V_1 \cup V_2$ is a balanced set of size $2\beta +2$, so there must be an edges (call it $e_1 := (b_1,b_2)$) connecting two of its vertices. 
Note that $e_1$ connects a vertex from $V_1$ to a vertex from $V_2$ (since the graph is bipartite). Suppose w.l.o.g. $b_1 \in V_1$ and $b_2 \in V_2$. 
Let $P':= C[b_1,b_2]$. Let $\mathcal{P}_3$ be the set of  paths from $\mathcal{P}_1$ contained in $P'$. Note that $|\mathcal{P}_3| \geq 2\beta + 2$. Let $V'_1$ be the set of start vertices of the first $\beta + 1$ paths in $\mathcal{P}_3$ and $V'_2$ the set of end vertices of the last $\beta + 1$ paths in $\mathcal{P}_3$. Observe that $V'_1 \cup V'_2$ is a balanced set of size $2\beta +2$, so there must be an edge (call it $e_2 := (w_1,w_2)$) connecting two of its vertices. Note that $e_2$ connects a vertex from $V'_1$ to a vertex from $V'_2$ (since the graph is bipartite). Suppose w.l.o.g. $w_1 \in V'_1$ and $w_2 \in V'_2$. Observe that if $e_2$ is blue then it forms a 0 biskip, so it must be red. Now observe that $C' := e_1 \cup e_2 \cup C[b_1,w_1] \cup C[b_2,w_2] $ is an $M$-alternating cycle with $0< w_M(C')\leq 2$ (resp. $-2\leq w_M(C')< 0$)  and the number of red (resp. blue) edges it contains is equal to the absolute value of its weight. 
\end{proof}

\subsection{BCPM in Bipartite Graphs}

In this subsection we prove that BCPM on bipartite graphs can be solved in polynomial time, i.e.\ we prove Theorem~\ref{th:FPTBCPM}.
We will first show that the following related problem on an integer-weighted directed graph can be solved in polynomial time, and then show how BCPM in bipartite graphs can be reduced to it.

\medskip

\noindent\vspace{5pt}\begin{boxedminipage}{\textwidth}
\textsc{Minimum Odd Cycle Problem (MOCP)}

\textbf{Input:} A digraph $D$ and a non-negative integral weight assignment $\vec{w}:A(D)\rightarrow \mathbb{N}_0$ to the arcs of $D$.

\textbf{Task:} Compute a directed cycle $C$ in $D$ with minimum total weight $\vec{w}(C)$ subject to $\vec{w}(C)$ being odd, or correctly conclude that no directed cycle $C$ in $D$ with odd weight exists.
\end{boxedminipage}

In the following, a \emph{walk} $Z$ in a digraph $D$ refers to an alternating sequence of vertices and arcs  $Z=v_1,e_1,v_2,\ldots,v_k,e_k,v_{k+1}$, where $e_i$ is an arc from $v_i$ to $v_{i+1}$ for $i=1,\ldots,k$. Repetitions of vertices or arcs are allowed. We refer to $k$ as the \emph{length} of $Z$, and call $Z$ a \emph{closed walk} if $v_{k+1}=v_1$. If $D$ is weighted by a function $\vec{w}:A(D)\rightarrow \{0,1\}$, we denote $\vec{w}(Z):=\sum_{i=1}^{k}{\vec{w}(v_i,v_{i+1})}$ for the total weight of arcs traversed by the walk (counted with multiplicities). The following is a simple yet crucial insight to solve the above problem. 

\begin{lemma}\label{lemma:walkstocycles}
Let $D$ be a digraph, $\vec{w}:A(D)\rightarrow \mathbb{N}_0$ a weight assignment. There is a deterministic algorithm that, given as input a closed walk $Z$ in $D$ with odd weight $\vec{w}(Z)$, outputs a directed cycle $C$ in $D$ such that $\vec{w}(C)$ is odd and $\vec{w}(C)\le \vec{w}(Z)$. The algorithm runs in time $\text{poly}(k)$, where $k$ is the length of $Z$.
\end{lemma}
\begin{proof}
The algorithm works as follows: Let $Z=v_1,e_1,v_2,\ldots,v_k,e_k,v_{k+1}=v_1$. Traverse the cyclic sequence $v_1,\ldots,v_k$ of vertices starting from $v_1$ until for the first time a vertex repeats. If no repetition occurs, $Z$ is a directed cycle and returning $Z$ does the job. 
Otherwise after $\text{poly}(k)$ steps we find two indices $1 \le i<j\le k$ such that $v_i=v_j$. Next we consider the walks $Z_1:=v_1,e_1,v_2,\ldots,e_{i-1},v_i=v_j,e_j,v_{j+1},\ldots,v_k,v_{k+1}=v_1$ and $Z_2:=v_i,e_{i},v_{i+1},\ldots,e_{j-1},v_j=v_i$. We compute the weights $\vec{w}(Z_1)$ and $\vec{w}(Z_2)$, and since $\vec{w}(Z_1)+\vec{w}(Z_2)=\vec{w}(Z)$ we can find $i \in \{1,2\}$ such that $\vec{w}(Z_i)$ is odd and $\vec{w}(Z_i)\le\vec{w}(Z)$. We then proceed by recursively calling the algorithm on the instance $Z_i$, which will yield a directed cycle $C$ in $D$ with $\vec{w}(C)$ odd and $\vec{w}(C)\le \vec{w}(Z_i)\le \vec{w}(Z)$, as desired. Since $Z_i$ is strictly shorter than $Z$, the algorithm described above in total can make at most $O(k)$ recursive calls and between two calls needs time $\text{poly}(k)$. Hence, the algorithm runs in time $\text{poly}(k)$. 
\end{proof}

Note that every directed cycle $C$ in a $\mathbb{N}_0$-weighted digraph $D$ of odd total weight contains an arc of odd weight. Thus, and using the previous lemma, we can see that the MOCP is polynomially equivalent to the following problem: 
For a fixed arc $e \in A(D)$ with odd weight $\vec{w}(e)$, among all closed walks using $e$ \emph{exactly once}, compute one of minimum odd total weight, or correctly conclude that no such odd weight walk exists. 

Indeed, given a solution to this problem, we can pick an arc $e$ of odd weight for which the returned solution is as small as possible, and then, if required, apply Lemma~\ref{lemma:walkstocycles} to turn the directed closed walk we found through $e$ into a directed cycle with at most the same odd total weight (which hence will be as small as possible). If for no arc $e$ of odd weight an odd weight closed walk was found, we can safely conclude that $D$ contains no odd weight directed cycle.  
We will accomplish this task in the next lemma, which then also completes the proof that the MOCP can be solved in polynomial time. 

\begin{lemma}
Let $D$ be a digraph, $\vec{w}:A(D)\rightarrow \mathbb{N}_0$, and $e \in A(D)$ with odd weight. There exists a deterministic polynomial-time algorithm that among all closed walks using $e$ exactly once, computes one of minimum odd total weight, or correctly concludes that no such closed walk exists.
\end{lemma}
\begin{proof}
Let $u$ and $v$ denote the starting- and endpoint of $e$, respectively. Note that since $\vec{w}(e)$ is odd the problem we want to solve is equivalent to computing a minimum even-weight directed $v$ to $u$-walk in $D-e$. 

Let $D',w'$ be another weighted digraph generated from $D-e=(V,A\setminus\{e\})$ and the arc-weights $\vec{w}(\cdot)$ (in polynomial time) as follows: $D'$ has vertex-set $V \times \{0,1\}$. Furthermore, for every arc $(x,y) \in A\setminus\{e\}$ with even weight $\vec{w}(x,y)$ we put two arcs, namely from $(x,0)$ to $(y,0)$ and $(x,1)$ to $(y,1)$ into $D'$, and assign weight $\vec{w}(x,y)$ to both of these arcs in $w'$. 
In contrast, for every arc $(x,y) \in A\setminus\{e\}$ with odd weight $\vec{w}(x,y)$ we put the two arcs, from $(x,0)$ to $(y,1)$ and $(x,1)$ to $(y,0)$ into $D'$, and assign weight $\vec{w}(x,y)$ to both of these arcs in $w'$. 

Using the above definitions, it is easy to check that there is a weight-preserving one-to-one correspondence between $v$ to $u$ walks in $D-e$ of even weight and $(v,0)$ to $(u,0)$ walks in $D'$. The problem in the statement of the lemma can thus be solved as follows: Generate $D'$ and $w'$ from $D$, $e$ and $\vec{w}$ in polynomial time. Then use one of the standard shortest weighted directed path algorithms (running in polynomial time) from $(v,0)$ to $(u,0)$ in the weighted digraph $D'$. If no directed path from $(v,0)$ to $(u,0)$ exists in $D'$, then no even weight $v$ to $u$-walk exists in $D-e$, and hence no odd weight walk in $D$ traverses $e$ exactly once. 

Otherwise, we obtain a minimum weight directed path $P$ from $(v,0)$ to $(u,0)$ in $D'$. Projecting vertices in $D'$ to their first coordinate yields a directed $v$ to $u$ walk in $D$ with the same minimum total even weight as $P$, and adding in $e$ yields a closed directed walk of smallest possible odd weight in $D$ using $e$ exactly once.
\end{proof}

Now that we have established a deterministic polynomial-time solution of the MOCP, we are ready to use it to solve the BCPM in bipartite graphs. 
\begin{proof}[Proof of Theorem~\ref{th:FPTBCPM}]
Let $G$ be bipartite input graph with bipartition into sets $A$ and $B$ and edges colored by red or blue. Using one of the polynomial algorithms for the minimum weight perfect matching problem in bipartite graphs, we first compute a perfect matching $M_0$ of $G$ with a minimum number of red matching edges. 
We first check whether $M_0$ contains at most $k$ red edges. If it contains more than $k$ red edges, then we can stop the algorithm and return that a perfect matching with at most $k$ red edges in $G$ does not exist. 
If $r(M_0) \le k$, we go on by checking whether $r(M_0) \equiv_2 k$, in which case we return $M_0$ as the solution to the BCPM instance given to us. 
So, in the remainder of the proof, we may assume that $r(M_0) \le k$ and $r(M_0) \equiv_2 k+1$.

Let $D$ be the directed graph which is an orientation of the edges of $G$ as follows: An edge $ab$ in $G$ with $a \in A, b\in B$ is oriented from $a$ to $b$ if $ab \in M_0$, and from $b$ to $a$ if $ab \notin M$. Note that a cycle in $G$ is an $M_0$-alternating cycle if and only if it is directed in $D$. 

As usual, let $w_0:E(G)\rightarrow \{0,1\}$ be defined by $w_0(e):=0$ if $e$ is a blue edge, $w_0(e):=1$ if $e$ is a red edge not in $M_0$ and $w_0(e):=-1$ if $e$ is a red edge in $M_0$. Note that for every other perfect matching $M$, we have $r(M)=r(M_0)+w_0(M \Delta M_0)$. This first of all implies, together with the minimality of $M_0$, that for every directed cycle $C$ in $D$, we have $w_0(C)=r(M_0 \Delta C)-r(M_0) \ge 0$. 

It further implies that a perfect matching $M$ in $G$ satisfies $r(M) \le k$ and $r(M) \equiv_2 k$ if and only if $w_0(M \Delta M_0)$ is odd and $w_0(M \Delta M_0) \le k-r(M_0)$. 

Suppose for a moment that such a matching $M$ exists. Then $M \Delta M_0$ corresponds to a disjoint union of cycles in $G$ which are directed in $D$, and $w_0(M \Delta M_0)$ being odd implies that at least one of these cycles, call it $C$, must also have an odd weight $w_0(C)$. Furthermore, all cycles in $M \Delta M_0$ are directed in $D$ and thus have non-negative weight, thus implying $w_0(C)\le w_0(M\Delta M_0)$. 

But then the perfect matching $M':=M \Delta C$ in $G$ has weight satisfying $r(M')=r(M_0)+w(C) \equiv_2 (k+1)+1\equiv_2 k$, and $r(M')=r(M_0)+w(C) \le r(M_0)+w_0(M\Delta M_0)=r(M)\le k$. Hence, we have proved that a perfect matching $M$ in $G$ satisfying $r(M) \le k$ and $r(M) \equiv_2 k$ exists if and only if there is such a matching with the additional property that $M \Delta M_0$ consists of a single directed cycle in $D$. 

Hence, finding such an $M$ is equivalent to finding a directed cycle $C$ in $D$ of odd total weight $w_0(C)$ such that in addition $w_0(C)\le k-r(M_0)$, or concluding that such a directed cycle does not exist. 

Note that in order to solve this problem it is not feasible to simply apply the polynomial-time solution to the MOCP to $D$ with the weight function $w_0$ immediately, since $w_0$ may not be non-negative. 

However, we can use a standard trick (described in the following) to transform $w_0$ into a non-negative integral weighting $\vec{w}$ of $D$ such that for every directed cycle $C$ in $D$, we have $w_0(C)=\vec{w}(C)$. 
To do so, consider an auxiliary weighted digraph $D^\ast$, obtained from $D$ by adding a new dominating source vertex $s$, i.e., $V(D^\ast)=V(D)\cup \{v^\ast\}$ and we add all the arcs $(s,v)$ with $v \in V(D)$ to $D^\ast$ and assign them weight $0$, while all original arcs retain their weight from $w_0$. Recall that every directed cycle $C$ in $D$ has non-negative weight $w_0(C)\ge 0$, and hence the same is true for $D^\ast$. We may thus use the algorithm of Bellman-Ford to compute, for every vertex $v \in V(D)$, the minimum weight of a directed walk in $D^\ast$ from $s$ to $v$. Denote this quantity by $p(v)$ for every $v \in V(D)$. We can now compute the modified weighting $\vec{w}$ of $D$ as follows. For every arc $e \in A(D)$ with start-vertex $u$ and end-vertex $v$, we set
$$\vec{w}(u,v):=w_0(u,v)+p(u)-p(v).$$
Using the above formula and a telescopic sum it is easy to see that $\vec{w}(C)=w_0(C)$ for every directed cycle $C$ in $D$, as desired. Furthermore, $\vec{w}$ is indeed non-negative: The definition of the potentials $p(\cdot)$ directly implies $p(v) \le p(u)+w_0(u,v)$ for every arc $(u,v)$ in $D$, and rearranging yields $w_0(u,v)+p(u)-p(v) \ge 0$. 

Having found the non-negative weighting $\vec{w}$, we can now apply the polynomial-time solution of the MOCP to $D$ with $\vec{w}$. This returns a directed cycle $C$ with smallest possible odd weight $w_0(C)=\vec{w}(C)$, or we conclude that no odd-weight cycle exists. In the latter case, we return that the BCPM in $G$ does not have a solution. In the first case, we check whether $w_0(C) \le k-r(M_0)$. If so, we compute the matching $M_0\Delta C$ and return this as the solution to the BCPM. Otherwise, we correctly conclude that BCPM in $G$ with the given edge-coloring has no solution. 
\end{proof}

\subsection{Main Theorem without Oracle Access}

\begin{proof}[Proof of \Cref{thm:approx-thm-alpha}]
The proof is very similar to that of \Cref{thm:main-thm-alpha}, with some minor modification.
First, instead of using CPM and BCPM to get $M$ and $M'$ for the input of the algorithm of \Cref{sec:algorithm}, we will simply let $M$ be a PM with a minimum number of red edges, and $M'$ a PM with a maximum number of red edges. Observe that such PMs can be computed in polynomial time by giving red edges positive or negative weights (to get $M$ and $M'$ respectively) and blue edges zero weights, and using any algorithm for minimum weight perfect matching that runs in deterministic polynomial time. Note that if any of the two matchings we are looking for did not exist, then the EM-instance would be a "No"-instance.
Now we apply the algorithm of \Cref{sec:algorithm} on the EM-instance with $M$ and $M'$ as input and consider a solution matching to be a PM that contains either $k-1$ or $k$ red edges.

Our goal is to again prove that if the EM-instance is a "YES" instance, then the following must be true:
\begin{itemize}
    \item[(a)] Phase 1 runs in polynomial time and outputs two PMs $M$ and $M'$ such that $r(M) \leq k \leq r(M')$ and $r(M') - r(M) \leq t$ (for $t = 16\cdot 4^{\alpha}$).
    \item[(b)] Phase 2 runs in polynomial time and either outputs a PM with $k-1$ or $k$ red edges (and the algorithm terminates) or a PM $M$ such that there exists a PM $M^*$ with $r(M^*) = k$ and $|E(M \Delta M^*)| \leq 2^{\alpha^{O(1)}}$ (for appropriately large constants).
    \item[(c)] If the algorithm did not terminate in Phase 2, then Phase 3 runs in time $f(\alpha)n^{O(1)}$ and outputs a PM with $k$ red edges.
\end{itemize}

It is easy to see that if all of the above items hold, then the algorithm runs in time $f(\alpha)n^{O(1)}$ and always outputs a PM with $k$ or $k-1$ red edges if one exists. Note that (a) and (c) again follow directly from \Cref{th:falphacloseparity} and \Cref{prop:smallsetcycles} respectively.

To prove (b) first observe that as long as neither $M$ nor $M'$ is a solution, all steps in phase 2 maintain the following invariants: $r(M) \leq k $ and $ r(M') \geq k-1$. To see this simply note that $r(M)$ and $r(M')$ can only change by $2$ every step. So in order for $r(M)$ to go above $k$ or $r(M')$ to go below $k-1$ they would need to pass by $k-1$ or $k$, at which point the algorithm terminates. Also observe that if any of the steps does not fail, then either $r(M') - r(M)$ decreases or $|E(M \Delta M')|$ decreases while $r(M') - r(M)$ remains unchanged. So if we consider as a measure of progress $r(M') - r(M)$ and $r(M') - r(M)$ ordered lexicographically (where progress is towards smaller values of the measure), then we always make progress (i.e. the measure strictly decreases). Note that $r(M') - r(M) \leq n$ and is always non-negative and the same holds for $|E(M \Delta M')|$. So the algorithm can perform at most $n^2$ iterations in phase 2. Since every iteration runs in polynomial time (this is true for steps (i) and (ii) by \Cref{prop:smallsetedges} and for step (iii) by \Cref{lem:boundedcycleweights}, \Cref{lem:notmanycycles} and \Cref{lem:notmanyredandblue}), we get that phase 2 runs in polynomial time.
Now observe that the algorithm only terminates in phase 2 if either $M$ or $M'$ is a solution (i.e. it has $k-1$ or $k$ red edges). So it remains to show that if the algorithm does not terminate in this phase then there exists a PM $M^*$ with $r(M^*) = k$ and $|E(M \Delta M^*)| \leq 2^{\alpha^{O(1)}}$. 
Observe that in case of non-termination, all the conditions of \Cref{lem:symdiffbound} are met: 
\begin{itemize}
    \item[(a)] $r(M) < k-1$, $r(M') > k$: follows from the invariants and $M$, $M'$ not being solutions.
    \item[(b)] $|w_M(M \Delta M')| \leq 256\cdot 4^{2\alpha}$: follows from $r(M') - r(M) \leq  16\cdot 4^{\alpha}$.
    \item[(c)] There is no PM $M_1$ such that $r(M_1) = r(M) + 2 $ and $|R(M \Delta M_1)| = 2$: follows from the failure of (i).
    \item[(d)] There is no PM $M_1'$ such that $r(M_1') = r(M') - 2$ and $|B(M' \Delta M_1')| = 2$: follows from the failure of (ii).
    \item[(e)] $M \Delta M'$ does not contain any 0-skip: follows from the failure of (iii).
    \item[(f)] The algorithms of \Cref{lem:boundedcycleweights}, \Cref{lem:notmanycycles} and \Cref{lem:notmanyredandblue} all fail to find a 0-skip-cycle set in $M \Delta M'$: : follows from the failure of (iii).
\end{itemize}
So by \Cref{lem:symdiffbound} we get the desired result.
\end{proof}